%% file: gnnrecs.tex
\documentclass[sigconf, natbib=false]{acmart}

\input{macros}      
\input{proj-macros}  
\input{acronyms}

\usepackage{listings}
\usepackage[utf8]{inputenc}
\usepackage[english]{babel}

\usepackage[mincrossrefs=99,style=ieee,maxcitenames=50,maxbibnames=99, minbibnames=99, mincitenames=4,useprefix=true,sorting=none]{biblatex}
\usepackage{tablefootnote}
\usepackage{multicol}
\graphicspath{{figures/}}

\setcopyright{rightsretained} 
\acmConference[KDD '25 MLoG-GenAI]{Knowledge Discovery and Data Mining Workshop on Machine Learning on Graphs in the Era of Generative Artificial Intelligence}{August 06, 2025}{Toronto, ON, Canada}
\acmYear{2025}
\acmDOI{}
\acmISBN{}

\begin{document}

\def\papertitle{Z-REx: Human-Interpretable GNN Explanations for Real Estate Recommendations}
\title{\papertitle}

\author{Kunal Mukherjee }
\email{kunal.mukherjee@utdallas.edu}
\affiliation{%
  \institution{The University of Texas at Dallas}
  \city{Richardson}
  \state{Texas}
  \country{USA}
}\authornote{This work was done while he was interning at Zillow Group, Inc., during summer '24.}
\author{Zachary Harrison}
\email{zacharyha@zillowgroup.com}
\affiliation{%
  \institution{Zillow Group, Inc}
  \city{Seattle}
  \state{Washington}
  \country{USA}
}
\author{Saeid Balaneshin}
\email{saeidb@zillowgroup.com}
\affiliation{%
  \institution{Zillow Group, Inc}
  \city{Seattle}
  \state{Washington}
  \country{USA}
}

%



\input{sections/abstract.tex}

\maketitle

\input{sections/introduction.tex}
\input{sections/background.tex}

\input{sections/problem.tex}
\input{sections/method.tex}
\input{sections/evaluation.tex}
\input{sections/conclusion.tex}

\printbibheading{}
\printbibliography[heading=none]

\input{sections/appendix}

\end{document}

%% file: macros.tex
\usepackage{caption}
\usepackage{subcaption}

\usepackage[framemethod=tikz]{mdframed}
\usepackage[most]{tcolorbox}
\usepackage[normalem]{ulem}
\usepackage{algorithm}
\usepackage{algpseudocode}
\usepackage{amsfonts}
\usepackage{amsmath}
\usepackage{arydshln}
\usepackage{balance}
\usepackage{blindtext}
\usepackage{booktabs}
\usepackage{cancel}
\usepackage{caption}[font={footnotesize}]
\usepackage{color,soul}
\usepackage{comment}
\usepackage{enumitem}
\usepackage{fontawesome}
\usepackage{graphicx}
\usepackage{hyperref,xcolor}
\usepackage{hyperref}
\hypersetup{hidelinks}
\usepackage{listings}
\usepackage{mathtools}
\usepackage{pifont}
\usepackage{tabularx}
\usepackage{threeparttable}
\usepackage{tikz}
\usepackage{xcolor,colortbl}
\usepackage{xspace}
\usepackage{wasysym}

\usepackage{makecell}
\usepackage{svg}

\definecolor{winered}{rgb}{0.7,0,0}
\definecolor{gray}{gray}{0.7}
\definecolor{darkpastelgreen}{rgb}{0.01, 0.75, 0.24}
\definecolor{cadmiumgreen}{rgb}{0.0, 0.42, 0.24}
\definecolor{brickred}{rgb}{0.8, 0.25, 0.33}
\definecolor{cornellred}{rgb}{0.7, 0.11, 0.11}
\definecolor{burgundy}{rgb}{0.5, 0.0, 0.13}
\definecolor{frenchblue}{rgb}{0.0, 0.45, 0.73}
\definecolor{light-gray}{gray}{0.92}
\definecolor{lightlight-gray}{gray}{0.97}
\definecolor{codegray}{gray}{0.90}
\definecolor{inputgray}{gray}{0.90}
\definecolor{darkgreen}{RGB}{40,125,40}

\hypersetup
{
colorlinks=true,
linkcolor=winered,
urlcolor={winered},
filecolor={winered},
citecolor={winered},
allcolors={winered}
}

\newcommand{\cmt}[1]{}

\newcommand{\homedir}{\raise.17ex\hbox{$\scriptstyle\sim$}}

\global\mdfdefinestyle{rtboxstyle}{%
linecolor=black,%
leftmargin=0cm,rightmargin=0cm,linewidth=0.5pt,
roundcorner=3,
skipbelow=0pt,backgroundcolor=lightlight-gray
}

\newcommand{\code}[1]{\texttt{#1}}

\mdfdefinestyle{MyFrame}{%
     roundcorner=1pt,
     innertopmargin=6pt,
     innerbottommargin=6pt,
     innerrightmargin=6pt,
     innerleftmargin=6pt,
    }

\graphicspath{{figure/}{figures/}}
\DeclareGraphicsExtensions{.pdf,.eps,.png,.jpg,.jpeg}

\long\def\comment#1{}

\renewcommand{\paragraph}[1]{\smallskip\noindent\emph{#1}\quad}
\def\Snospace~{\S{}}

\newcommand{\heading}[1]{{\vspace{2pt}\noindent\bf{#1}}} 

\newcommand{\eg}{{\it e.g.,~}}
\newcommand{\ie}{{\it i.e.,~}}

\newcommand{\nm}[1]{{\it (#1)\xspace}} 

\newcommand{\optional}[1]{}

%% file: proj-macros.tex
\def\pname{{\textsc{ZiGNN}}\xspace} 
\def\pnameexp{{\textsc{Z-REx}}\xspace} 
\usepackage{fp}

\newif\ifbiblatex
\biblatextrue

\newif\ifcomments
\commentstrue

\ifcomments
\usepackage{todonotes}
	\newcommand{\TODO}[1]{\textcolor{red}{TODO: #1}}
	\newcommand{\wc}[1]{}
	\newcommand{\kunal}[1]{\textcolor{orange}{[{\bf KUNAL:} #1]}}
	\newcommand{\zach}[1]{\textcolor{blue}{[{\bf ZACH:} #1]}}
	\newcommand{\fix}[1]{\textcolor{red}{\hl{#1}}}
	\newcommand{\corr}[2]{\sout{#1} \hl{#2}}
\else
	\newcommand{\TODO}[1]{}
	\newcommand{\wc}[1]{}
	\newcommand{\kunal}[1]{}
	\newcommand{\zach}[1]{}
	\newcommand{\fix}[1]{}
	\newcommand{\corr}[2]{}
\fi

\def\gnne{GNNExplainer\xspace}%
\def\subx{SubgraphX\xspace}%
\def\pge{PGExplainer\xspace}%
\def\plink{PaGE-Link\xspace}%

%% file: acronyms.tex
\usepackage[acronym]{glossaries}

\newacronym{hdl}{HDL}{High-level Dynamic Language}

\newacronym{ml}{ML}{Machine Learning}
\newcommand{\ml}{\gls*{ml}\xspace}

\newacronym{lp}{LP}{Link Prediction}
\newcommand{\lp}{\gls*{lp}\xspace}

\newacronym{ai}{AI}{Artificial Intelligence}

\newacronym{nn}{NN}{Neural Network}

\newacronym{gnn}{GNN}{Graph Neural Network}
\newcommand{\gnn}{\gls*{gnn}\xspace}
\newcommand{\gnns}{\glspl*{gnn}\xspace}

\newacronym{dnn}{DNN}{Deep Neural Network}

\newacronym{llm}{LLM}{Large Language Model}

\newacronym{mlm}{MLM}{Masked Language Model}

\newacronym{rnn}{RNN}{Recurrent Neural Network}

\newacronym{gan}{GAN}{Generative Adversarial Network}

\newacronym{nlp}{NLP}{Natural Language Processing}

\newacronym{pl}{PL}{Programming Language}

\newacronym{abi}{ABI}{Abstract Binary Interface}

\newacronym{vm}{VM}{Virtual Machine}

\newacronym{bert}{BERT}{Bidirectional Encoder Representations from Transformers}

\newacronym{csn}{CSN}{CodeSearchNet}

\newacronym{pypi}{PyPI}{Python Package Index}

\newacronym{sota}{SOTA}{State-of-The-Art}
\newcommand{\sota}{\gls*{sota}\xspace}

\newacronym{cdm}{CDM}{Common Data Model}

\newacronym{dgl}{DGL}{Deep Graph Library}
\newcommand{\dgl}{\gls*{dgl}\xspace}

\newacronym{tc}{TC}{Transparent Computing}

\newacronym{dt}{DT}{Decision Tree}

\newacronym{apt}{APT}{Advanced Persistent Threat}

\newacronym{cve}{CVE}{Common Vulnerabilities and Exposures}

\newacronym{ids}{IDS}{\emph{Intrusion Detection System}}

\newacronym{pids}{PIDS}{\emph{Provenance-based Intrusion Detection System}}

\newacronym{av}{AV}{Anti-Virus}

\newacronym{edr}{EDR}{End-host Detection and Response}

\newacronym{poi}{POI}{Point-Of-Interest}

\newacronym{vlan}{VLAN}{Virtual LAN}

\newacronym{ctwo}{C2}{Command and Control}

\newacronym{dll}{DLL}{Dynamic Link Library}

%% file: sections/abstract.tex
\begin{abstract}\label{sec:abstract}

    Transparency and interpretability are crucial for enhancing customer confidence and user engagement, especially when dealing with black-box \ml-based recommendation systems. Modern recommendation systems leverage \gnn due to their ability to produce high-quality recommendations in terms of both relevance and diversity. Therefore, the explainability of \gnn is especially important for \lp tasks since recommending relevant items can be viewed as predicting links between users and items. \gnn explainability has been a well-studied field, but existing methods primarily focus on node or graph-level tasks, leaving a gap in \lp explanation techniques.

    This work introduces \pnameexp, a \gnn explanation framework designed explicitly for heterogeneous link prediction tasks. \pnameexp utilizes structural and attribute perturbation to identify critical substructures and important features while reducing the search space by leveraging domain-specific knowledge. In our experimentation, we show the efficacy of \pnameexp in generating contextually relevant and human-interpretable explanations for \pname, a \gnn-based recommendation engine, using a real-world real-estate dataset from \underline{Zi}llow Group, Inc. We compare against \sota \gnn explainers to show \pnameexp outperforms them by 61\% in Fidelity metric by producing superior human-interpretable explanations.

\end{abstract}

%% file: sections/introduction.tex
\section{Introduction}\label{sec:introduction}

As user interactions in housing marketplaces grow in volume and complexity, personalized recommendation systems are becoming indispensable for improving user experience and satisfaction. Recent advancements in data modeling and recommendation algorithms enable the creation of detailed interaction graphs, that capture nuanced relationships between users and items (\eg property listings and cities). Leveraging the inherent structure of these graphs provides fine-grained insights into user behaviors and preferences, forming a strong foundation for advanced recommendation systems. Building on these insights, various learning-based recommendation systems have been developed and deployed to effectively utilize this data, significantly improving personalization and engagement rates across multiple domains.

In particular, \gnns have emerged as a powerful tool for modeling relational data in recommendation systems. \gnns leverage graph-structured data, where nodes represent entities (such as users, listings, and cities) and edges denote interactions (such as user views, saves, and tours). Through a message-passing mechanism, \gnns aggregate information from neighboring nodes and edges, enabling them to capture complex, multi-hop relationships within the graph. This capability makes \gnns particularly effective for personalized recommendations, as they can identify nuanced patterns in user preferences and item associations.

While \gnn-based recommendation models achieve impressive accuracy, their black-box nature poses significant challenges regarding trust and transparency. Users and system administrators often require explanations for recommendations, particularly in high-value transactions or critical decision-making scenarios. Current \gnn explanation techniques focus on node and graph-level tasks whereas recommendation systems model the recommendation problem as Link Prediction (LP) tasks, as recommending relevant items to users can be framed as predicting links between users and relevant items. Therefore, there is a notable gap in the literature concerning the explanation of \lp tasks. 


Existing \gnn explanation techniques~\cite{ying2019gnnexplainer, luo2020parameterized, yuan2021explainability} primarily focus on generating explanations by either learning a mask to select an edge-induced subgraph or searching for the most informative subgraph. While these methods are effective for node-level and graph-level tasks, they face significant challenges in link-level prediction contexts. For instance, approaches like \cite{ying2019gnnexplainer, luo2020parameterized} often yield disconnected edges or subgraphs, making the resulting explanations challenging to interpret about the predicted link. Furthermore, \cite{yuan2021explainability} suffers from scalability issues as the graph size increases exponentially, leading to computational challenges due to the combinatorial explosion of possible subgraphs. This is particularly problematic since enumerating all subgraph types requires checking for isomorphism, which results in exponential time complexity as the graph’s vertices and edges grow. Customizing these techniques for sparse datasets and heterogeneous graphs introduces further difficulties. Existing methods emphasize dense local subsets of the dataset and are primarily designed for homogeneous graphs, limiting their applicability to more sparse and diverse graph structures.


Explaining \gnns for \lp introduces three unique challenges (also noted by ~\cite{zhang2023page}): \nm{1} accurate interpretation of substructures in the presence of sparse relations, \nm{2} scalability, and \nm{3} heterogeneity. To address these challenges, we formulate \lp explanation as an instance-level, post-hoc task that generates interpretable ground-truth aware explanations by identifying the most important feature subset and critical subgraphs. Our approach offers both interpretable and scalable explanations while also accounting for the heterogeneity of real-world recommendation systems. This provides a critical advancement in improving the transparency and trustworthiness of \gnn-based recommendation systems.

We propose \pnameexp, an instance-level \gnn-based recommendation explanation framework for \lp tasks using entire heterogeneous graphs. \pnameexp leverages both graph structures and domain knowledge to identify critical structural relationships to provide explanations with better contextual relevance and alignment with ground truth. Our approach involves a two-step collaborative process where we perturb: \nm{1} the features to identify the most important subset, and \nm{2} the graph to determine the key subgraphs.

Our study emphasizes the importance of decoupling the decision-making process using both structural and entire feature sets, and instead focuses on using a subset of features and subgraphs to enhance the interpretability of recommendations. This allows us to create context-aware explanations that align closely with the interaction patterns found in recommendation systems. We stress that our framework, \pnameexp, is designed to provide interpretable insights into \gnn decisions, not to improve recommendation accuracy. 

To demonstrate the effectiveness of \pnameexp, we measure the quality of explanation for recommendations made by \pname, a \gnn-based recommendation engine for heterogeneous interaction graphs. We use a real-world real-estate dataset from Zillow Group, Inc. The dataset is collected from a large-scale recommendation platform, where user interactions are logged across a diverse set of listings and cities over an extended period. Finally, we compare it against \sota \gnn explainers, evaluating its ability to provide relevant explanations in recommendation tasks.  Our evaluation shows that \pnameexp outperforms existing explainers in terms of providing actionable and interpretable insights for both users and system administrators. This work highlights the potential for developing domain-specific features to close the gap between high-performing recommendation systems and their need for transparency.

In summary, the key contributions of our work are:

\begin{itemize}[noitemsep,topsep=1pt] 
    \item To the best of our knowledge, this is the first end-to-end framework that integrates recommendation and instance-level explanation using whole graph structure without disintegrating into paths or subgraphs.
    \item We conduct an extensive analysis of the real-world housing market to uncover structural heuristics that reduce the explanation search space and improve the relevance of recommendation explanations.
    \item We propose a collaborative approach combining feature and structural perturbation, which enhances both the accuracy and diversity of recommendations and is validated through a real-world case study.
    \item \pnameexp outperforms \plink by 29\%, \gnne by 85\%, and \subx by 70\% in the Fidelity metric, demonstrating its superior explanation accuracy.
\end{itemize}

%% file: sections/background.tex
\section{Background and Related Works}\label{sec:background}


\heading{\gnn-based Recommendation Framework.}
Graph Neural Networks (GNNs) have emerged as a powerful approach for enhancing recommender systems. Over the past few years, several studies have applied \gnns directly to user-item bipartite graphs, yielding significant improvements in both effectiveness and efficiency \cite{wang2023collaboration, mukherjee2023explaining, chen2020revisiting,sun2020neighbor,wang2019neural}. One common challenge in this approach lies in capturing higher-order connectivity between nodes. Multi-GCCF \cite{sun2019multi} and DGCF \cite{liu2020deoscillated} address this by introducing artificial edges that connect two-hop neighbors (\eg user-user and item-item graphs) to introduce proximity information into the user-item interaction.

Node representations are computed layer-by-layer in GNNs, where the overall user and item representations are critical for downstream tasks like recommendation prediction. The most common practice is to adopt the final-layer embeddings as the ultimate representations \cite{li2019hierarchical,ying2018graph}. 
For a comprehensive survey of GNN-based recommender systems, readers are referred to \cite{wu2022graph}.

NGCF \cite{wang2019neural} enhances feature interactions between users and items using an element-wise product operation. NIA-GCN \cite{sun2020neighbor}, on the other hand, introduces pairwise neighborhood aggregation to better capture neighbor relationships. Inspired by GraphSAGE \cite{hamilton2017inductive}, \cite{li2019hierarchical,sun2019multi,ying2018graph} utilize a concatenation operation followed by nonlinear transformations to update node representations. Conversely, LightGCN \cite{he2020lightgcn} and LR-GCCF \cite{chen2020revisiting} simplify the aggregation by removing non-linearities, which enhances both performance and computational efficiency.

\heading{\gnn-based Explainers.}
Recent research in \gnn explainers \cite{ying2019gnnexplainer, luo2020parameterized, yuan2021explainability} has advanced in identifying key nodes, edges, or subgraphs in \gnns. They are categorized into white-box and black-box explainers. White-box methods, \eg \gnne \cite{ying2019gnnexplainer} and \pge \cite{luo2020parameterized}, access \gnn internals, including model weights and gradients. Conversely, black-box methods like \subx \cite{yuan2021explainability} operate on model inputs and outputs, reducing coupling between the explanation framework and model architecture. \gnn explainers encounter exponentially increasing computation time with graph size growth, hindering the interpretability in real-world graphs. 

\heading{\gnn-based Recommendation Explainers.}
The rise of \gnn-based recommender systems has created a pressing need for explainability in these recommendation systems~\cite{zhang2020explainable}. Explainable recommender systems aim to not only deliver accurate predictions but also provide transparent and persuasive justifications for recommendations~\cite{lyu2022knowledge,sinha2002role,wang2022multi,zhang2020explainable}.
Prior work on explainable recommender systems adopted these strategies~\cite{zhang2020explainable} of designing intrinsically explainable models with interpretable logic~\cite{zhang2014explicit, he2015trirank} and using post hoc models that generate explanations for the predictions of black-box models~\cite{peake2018explanation}. However, these methods face two key challenges: \nm{1} representing explainable information often requires node attributes and influential subgraphs identification, and \nm{2} reasoning for recommendations relies on domain knowledge~\cite{zhang2020explainable}.

Several explainable AI (XAI) approaches have been proposed, focusing on node or graph classification tasks \cite{duval2021graphsvx,lin2021generative,luo2020parameterized,pope2019explainability}. These methods commonly provide factual explanations in the form of subgraphs deemed relevant for a particular prediction analogous to feature-based XAI methods like LIME \cite{ribeiro2016should} and SHAP \cite{lundberg2017unified}.
~\cite{zhang2023page} was the first work to do a path-based graph neural network explanation for heterogeneous link prediction tasks with the primary limitation that the explanation depends on the ego-graph constructed around the explanation node. Therefore, the explanation technique cannot provide relevant explanations for larger graphs with multiple node attributes (\eg ego-graph size increases exponentially) and large graph diameter. Our approach bridges the gap between general XAI solutions and the unique need for explainable recommendations, addressing transparency challenges and the complexity of graph-structured user-item relationships.

%% file: sections/problem.tex




\section{Problem Statement}\label{sec:problem-statement}

In this section, we formally define the problem of providing interpretable explanations for the link prediction task in \gnn-based recommendation systems using whole heterogeneous graphs. The key challenge is to provide human-interpretable insights that align with the ground-truth so that user confidence and user trust is confirmed. Therefore, we aim to identify the most important feature subset and critical subgraphs used for recommendation by a \gnn-based recommendation engine using a real-world heterogeneous real-estate dataset (\eg Zillow Group, Inc).

\heading{Recommendation Task.} The \textit{primary task} of the recommendation system discussed here is to predict the likelihood of a link between a user \( u \in \mathcal{V}_u \) and a city \( l \in \mathcal{V}_c \) based on observed interactions and the graph structure. This is formalized as a link prediction problem: $\hat{y}_{uc} = f_{\theta} (u, c, \mathcal{G})$, where \( f_{\theta} \) is the link prediction model parameterized by \( \theta \), and \( \hat{y}_{uc} \) is the predicted probability of an interaction between user \( u \) and city \( c \).


\section{Preliminaries}

We outline the construction of interaction graphs used for recommendation and define the ground-truth data leveraged for training and evaluating our models.

\begin{figure}[h!]
	\centering
	\includegraphics[width=0.9\linewidth]{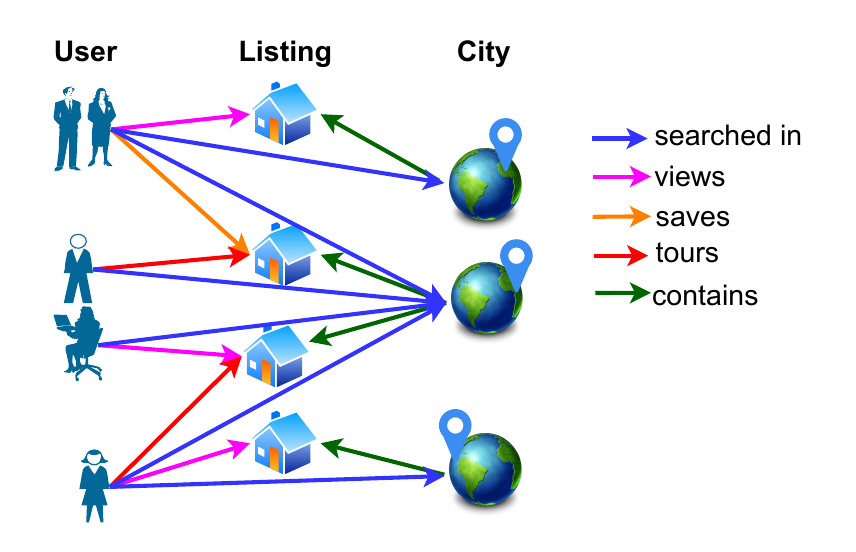}
	\caption{Interaction graph from a real-estate dataset.}
	\label{fig:igraph}
\end{figure}

\input{tables/edge-attr.tex}

\subsection{Interaction Graph}\label{sec:inter-graph}

The user interaction data in real-estate is inherently relational, comprising of entities such as users, listings, and cities, along with various interaction types like \code{views}, \code{saves}, and \code{tours} as shown in \autoref{fig:igraph} and listed in \autoref{tab:edge-attr} (more details in \autoref{sec:graph-details}). We focus primarily on three entities in this study since location is the most important factor in the home-buying process~\cite{harrison2022taxonomic}. All searches on Zillow Group, Inc website~\cite{zillowwebsite} correspond to a location, and we consider the city to be one of the entities, in addition to the interactions between users and listings. Since every listing belongs to a city, a special \code{contains} edge is added between the listing and the city. These entities and interactions are best modeled as a heterogeneous interaction graph \( \mathcal{G} = (\mathcal{V}, \mathcal{E}) \), where:
\begin{itemize}
    \item \( \mathcal{V} \) represents the set of nodes (\eg user, listing, and city)
    \item \( \mathcal{E} \subseteq \mathcal{V} \times \mathcal{V} \) denotes the set of edges that encode interactions (\eg user $\rightarrow$ \code{views} $\rightarrow$ listing, city $\rightarrow$ \code{contains} $\rightarrow$ listing).
\end{itemize}
The construction of the interaction graph \( \mathcal{G} \) involves aggregating data from multiple sources, such as user behavior logs, listing metadata, and geographical information. Each edge \( e \in \mathcal{E} \) is associated with a type \( \tau(e) \), representing the nature of the relationship between two nodes. Formally, we define a heterogeneous interaction graph as:
$\mathcal{G} = \left( \mathcal{V}_u \cup \mathcal{V}_l \cup \mathcal{V}_c, \mathcal{E}_{ul} \cup \mathcal{E}_{cl} \cup \mathcal{E}_{uc} \right)$, where \( \mathcal{V}_u \), \( \mathcal{V}_l \), and \( \mathcal{V}_c \) represent the sets of user, listing, and city nodes, respectively. \( \mathcal{E}_{ul} \) corresponds to interactions between users and listings, \( \mathcal{E}_{cl} \) captures relationships between cities and listings, and \( \mathcal{E}_{uc} \) captures relationships between users and cities. 

\subsection{Ground Truth}

The ground truth for training and evaluating is derived from historical user interactions with listings on the platform. Specifically, the interaction logs provide labels for whether a user \( u \) has engaged with a listing \( l \) (\eg viewed, saved, and toured), resulting in positive examples for link prediction. We infer the user to city interaction based on the listing interaction as listings are all associated with a city. 

\begin{figure}[h!]
	\centering
	\includegraphics[width=0.9\linewidth]{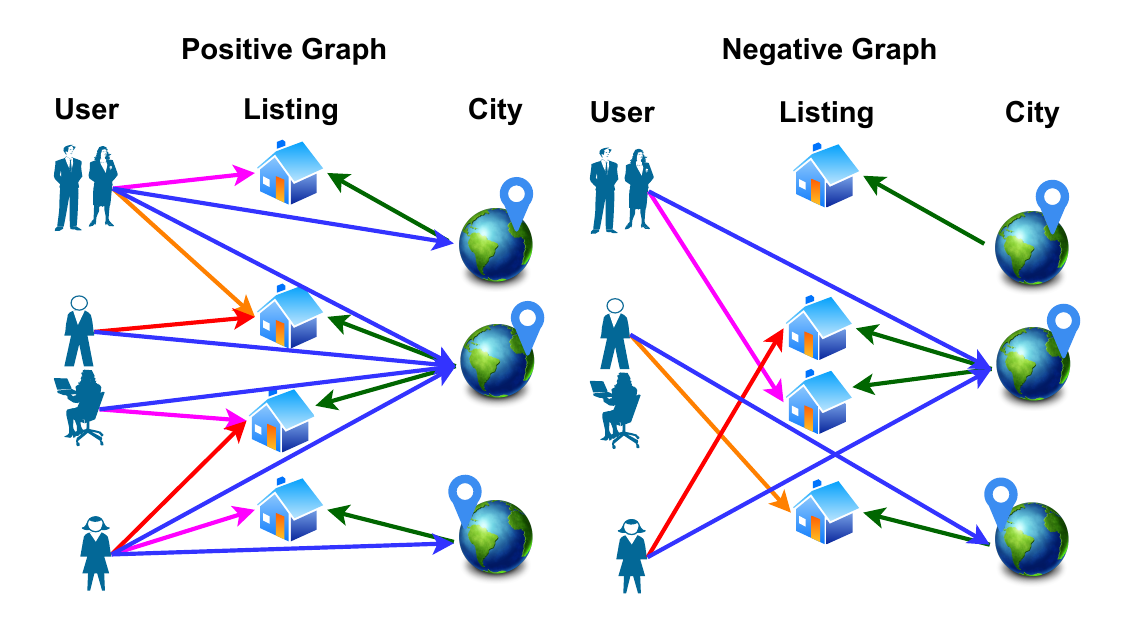}
	\caption{Negative graph construction from positive graph.}
	\label{fig:neg}
\end{figure}

Negative examples are defined by selecting edges between users and cities (or listing) where no interaction has been recorded, assuming that non-interaction implies a lack of interest. These are weak negative edges, as there is no explicit evidence that the user dislikes the item. However, for our use case in real-estate, we argue that a user's lack of interaction and interest are closely aligned, given the context where the items are either listings or cities. Thus, incorporating weak negative edges to create a negative graph helps the recommendation model learn finer distinctions about the city (or listing) preferences and understand why users do not interact with them. A negative graph is generated by sampling negative edges from user-listing and user-city pairs as shown in \autoref{fig:neg}. 

We generate ground truth explanations by identifying the subgraphs most relevant for each prediction and comparing their node features to identify their similarity. These subgraphs serve as interpretable justifications for why a specific user \( u \) is likely to engage with listing \( l \), thus providing a transparent recommendation system.

Let, \( \mathcal{Y} \subseteq \mathcal{V}_u \times \mathcal{V}_l \) be the set of observed interactions (positive examples), and \( \mathcal{Y'} \subseteq \mathcal{V}_u \times \mathcal{V}_l \) be set of sampled negative examples. The training set \( \mathcal{T} \) is constructed as $\mathcal{T} = \{ (u, l, y_{ul}) \mid (u, l) \in \mathcal{Y} \cup \mathcal{Y'}$,
where \( y_{ul} = 1 \) if \( (u, l) \in \mathcal{Y} \) and \( y_{ul} = 0 \) if \( (u, l) \in \mathcal{Y'} \).

%% file: tables/edge-attr.tex
\begin{table}[ht]
    \centering
    \resizebox{0.7\columnwidth}{!}{%
    \begin{tabular}{@{}lcc@{}}
    \toprule
    \textbf{\makecell{Src Node \\ Type}} & \textbf{\makecell{Dst Node \\ Type}} & \textbf{\makecell{Edge \\ Type}} \\ \midrule
    User & Listing & \code{views}, \code{saves}, and \code{tours} \\
    User & City & \code{searched in} \\
    City & Listing & \code{contains} \\
    \bottomrule
    \end{tabular}%
    }
    \caption{Description of relationships in the heterogeneous graph.}
    \label{tab:edge-attr}
\end{table}

%% file: sections/method.tex

\begin{figure*}[h!]
	\centering
	\includegraphics[width=0.8\linewidth]{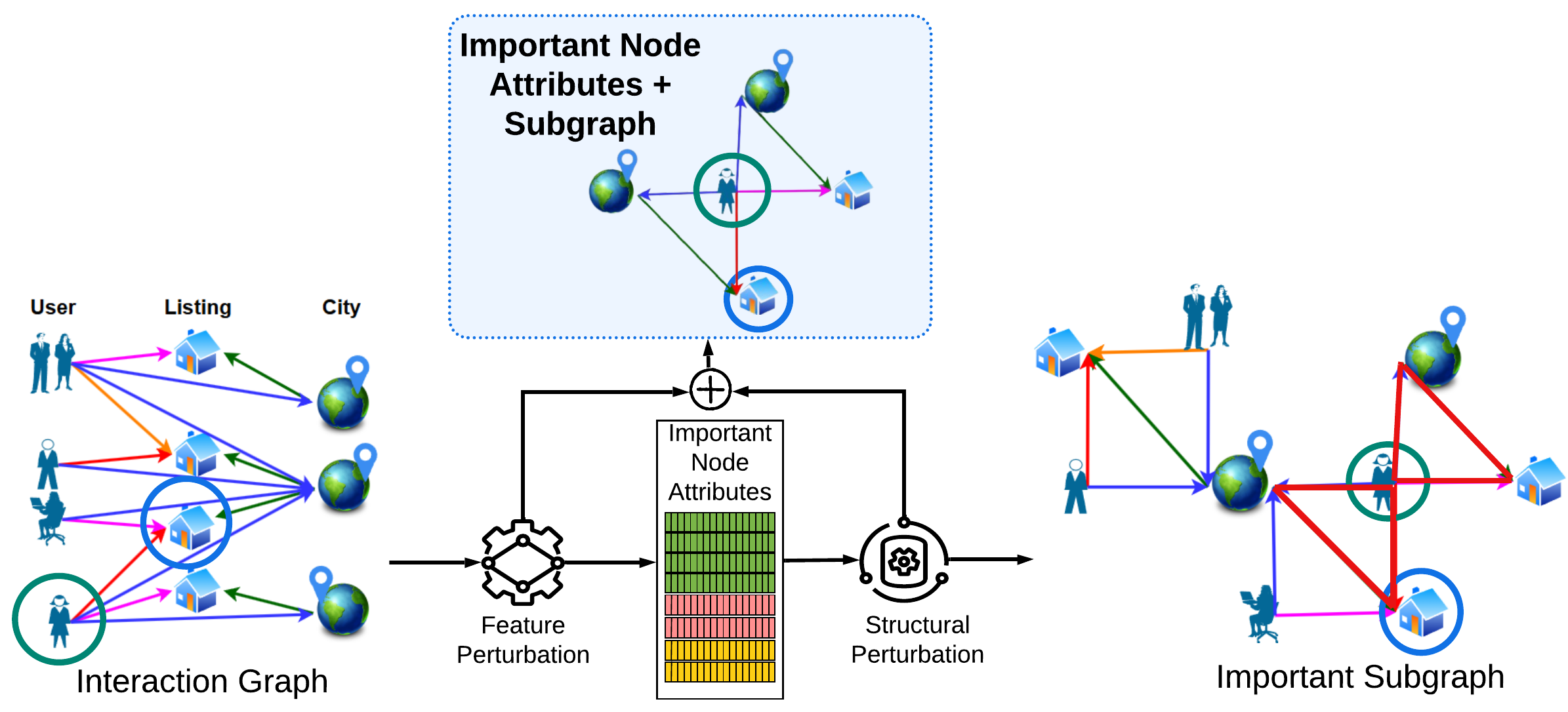}
	\caption{\pnameexp overview.}
	\label{fig:overview}
\end{figure*}

\section{\pnameexp Overview}\label{sec:design}





The workflow of \pnameexp is illustrated in \autoref{fig:overview} and algorithm is described in \autoref{alg:zrex} where \pnameexp first performs feature perturbation to find the important features and then performs graph structural perturbations to identify the important subgraph structures. We \textit{focus} on user-city due to the problem definition in \autoref{sec:problem-statement}, but this method works for any relationship. 

\subsection{Feature Perturbation.}

To interpret the \gnn-based recommendation system, we first analyze the influence of target node features using feature perturbation. Consider a user $u$ and their recommended city $c_t$ with their embeddings, $h_u$ and $h_{c_t}$, where cosine similarity represents the predicted affinity score that the \gnn uses to rank city $c_t$ for user $u$:

\begin{equation}
    \text{sim}(\mathbf{h}_u, \mathbf{h}_{c_t}) = \frac{\mathbf{h}_u \cdot \mathbf{h}_{c_t}}{\|\mathbf{h}_u\| \|\mathbf{h}_{c_t}\|}.
\end{equation}

The feature perturbation involves the following steps:
\begin{enumerate}    
    \item \textbf{Feature Perturbation:} 
    Perturb the target city's features $\mathbf{x}_{c_t}$ one by one by zeroing them out, and the indices of the features to be zeroed out are in $\mathcal{F}$. The perturbed features $\bar{\mathbf{x}}_{c_t}$ are defined as:
    \begin{equation}
        \bar{\mathbf{x}}_{c_t}[i] = 
        \begin{cases}
          0, & \text{if } i \in \mathcal{F}, \\
          \mathbf{x}_{c_t}[i], & \text{otherwise}.
        \end{cases}
    \end{equation}

    \item \textbf{Compare Performances:} 
    Compare the change in nDCG@K $\Delta \mathrm{nDCG}(\mathcal{F})$ due to the perturbed features $\bar{\mathbf{x}}_{c_t}$:
    \begin{equation}
        \Delta \mathrm{nDCG@K}(\mathcal{F}) 
        = \mathrm{nDCG@K}\bigl(\bar{\mathbf{x}}_{c_t}\bigr) 
        \;-\; \mathrm{nDCG@K}\bigl(\mathbf{x}_{c_t}\bigr).
    \end{equation}

\end{enumerate}

The \pname is re-evaluated using the perturbed features, and the resulting performance degradation is measured using the ranking metric of nDCG@K (Normalized Discounted Cumulative Gain). A significant drop in the metric indicates the importance of the perturbed feature for the recommendation.

\subsection{Structural Perturbation.}\label{sec:struc-pertb}

While feature perturbation focuses on node-level characteristics, structural perturbation analyzes the impact of graph topology on the model's predictions. Thus, accounting for the whole graph context in each recommendation explanation. Please note that structural perturbation happens with only the subset of features identified in the previous step. Feature perturbation identifies node attributes that have a significant influence on the recommendation outcomes, while structural perturbation uncovers graph edges and relationships that are critical to the model's predictions. 

Specifically, we study how the presence or absence of edges in the graph influences the recommendation similarity. The procedure is as follows:

\begin{enumerate}
    \item \textbf{Graph Transformation}: From the heterogeneous graph, a user-city graph $\mathcal{G}_h$ is created by collapsing all user-city relationships and removing intermediate nodes (e.g., listings). A $k$-hop subgraph $\mathcal{G}_u^k$ centered around a target user $u$ is then extracted, focusing on both direct and indirect relationships with city nodes.
    
    \item \textbf{Identify Co-clicked Cities}: For the subgraph $\mathcal{G}_u^k$, we identify pairs of cities $(c_i, c_j)$ that share a common predecessor user $u_p$. These pairs are added as new edges, representing co-click relationships, and their contributions to the model predictions are evaluated. By \textit{co-clicked cities}, we mean cities in which the current user clicked listings, as well as cities other users clicked listings in, where they share at least one common city with the current user. A co-clicked city indicates user groups with similar preferences.
    
    \item \textbf{Edge Removal and Similarity Change}: To assess the importance of structural connections, we iteratively remove identified edges and recompute the similarity between the user embedding $\mathbf{h}_u$ and the target city embedding $\mathbf{h}_{c'_t}$. The change in similarity $\Delta \text{sim}$ after edge removal is defined as:
    
    \begin{equation}
        \Delta \text{sim} = \text{sim}(\mathbf{h}_u, \mathbf{h}_{c'_t}) - \text{sim}(\mathbf{h}_u, \mathbf{h}_{c_t}).
    \end{equation}
    
\end{enumerate}

Edges with the highest absolute $\Delta \text{sim}$ values are identified as critical contributors to the recommendation. These edges represent strong graph relationships that drive user preferences for specific cities. The hyperparameters that influence the structural perturbations are: \nm{1} $k$ (hop distance) as it limits the number of edges perturbed to efficiently evaluate while focusing on the most impactful connections and increasing $k$ captures more indirect relationships but may introduce noise, and \nm{2} edge removal strategy as prioritizing edges based on shared predecessors ensures that only influential connections are analyzed. 

\subsection{Graph Size vs. Resource Overhead}

We provide real estate-specific recommendation explanations by leveraging the insight that co-clicked cities have a significant impact on recommendations. Co-clicked cities restrict the search space for structural perturbations, which in turn lead to finding the optimal number of edges required to change the recommendations. We acknowledge that iterative perturbation can be a resource-intensive process. But, our experiments show that real estate interaction is sparse in nature (as seen in \autoref{tab:interactions}), since people view the most number of listings, but only save a few, and tour even fewer listings. Therefore, the interaction graph from the real-world data is also sparse (as seen in the 30-days sub-section of \autoref{tab:interactions}), so large-scale real-world deployments are feasible.

As seen in \autoref{tab:runtime} in evalaution, training one epoch on a 30-day interaction graph containing 29.8M edges requires 10.89 sec, only an order of magnitude slower than the 0.88 sec required for a 3-day snapshot containing 1.2M edges, confirming that additional temporal coverage does not explode computation and inference time stays almost the same. The results demonstrate that \pnameexp delivers timely and faithful explanations at a production scale.

\subsection{Generalizability of \pnameexp}

To adapt the \pnameexp for other recommendation tasks, the edge perturbation step needs modification since for real estate, we are using domain knowledge and are focusing on co-clicked edges. For other domains, these edges might not exist, or the assumption we made might not be accurate. Since the structure influences the \gnn's message passing step, we think that for other domains, we can translate the insights to structural constraints, we can update \autoref{sec:struc-pertb} Structural Perturbation, and decrease the search space from all the potential edges to a smaller list of potentially important edges.

%% file: sections/evaluation.tex
\section{Evaluation}\label{sec:evaluation}

To comprehensively evaluate, \pnameexp we investigate the following research questions using the methodology~\autoref{sec:method} and dataset~\autoref{sec:data}.

\begin{itemize}[noitemsep,topsep=1pt]
	\item{\bf RQ1: Recommendation Accuracy.} Can \pname recommend relevant regions? (\autoref{sec:recs-acc})?
	\item{\bf RQ2: Explanation Accuracy.} Can \pnameexp explain \pname's recommendations ? (\autoref{sec:exp-acc})?
	\item{\bf RQ3: Comparison with \sota \gnn Explainers.} How do the explanations of \pname compare against those of \sota \gnn explainers (\gnne\cite{ying2019gnnexplainer}, \pge\cite{luo2020parameterized}, and \subx\cite{yuan2021explainability}) (\autoref{sec:sota-comp})?
\end{itemize}

\subsection{Methodology}\label{sec:method}

The data processing of \pname is described in \autoref{alg:preprocess}, which begins by extracting features containing user activities (\eg \code{views} and \code{saves}), user and listing attributes, and their geographical regions. During preprocessing, missing values are handled systematically to ensure data integrity. Outliers, which can distort analysis and model training, are addressed by replacing extreme values (detected using Z-scores) with the column mean. These steps ensure a clean and reliable dataset. Additionally, numerical features are normalized using z-score normalization, creating a consistent data representation. These preprocessing steps create a robust and standardized dataset, ready for graph-based modeling in the recommendation system.

After preprocessing, the user, listing, and city data are used to construct a heterogeneous graph. Each entity (\eg user, listing, and city) is represented as a unique node type, and edges capture interactions between these nodes. For instance, user-to-listing edges represent activities such as \code{views} or \code{saves}, while city-to-listing edges denote listings contained within a city. 
Node features are assigned and this comprehensive graph representation forms the backbone of the recommendation system, enabling the \pname to understand complex relationships.

The training pipeline first constructs the negative graph to ensure a balanced training dataset by mimicking real-world scenarios where users do not interact with all items. 
During evaluation, the  \pname's effectiveness was tested using node embeddings derived from the trained model. These embeddings were normalized to enable cosine similarity-based retrieval, crucial for recommendation tasks. For specific canonical edge types (\eg 'user', \code{views}, 'city'), embeddings were computed separately for source (\eg user) and destination nodes (\eg city). The recommendation performance of \pname is measured using nDCG@K, which evaluates the ranking quality of recommendations and prioritizes highly relevant items. Metrics like nDCG@K was used to evaluate recommendation quality, providing quantitative insights.

\heading{Metric.} \pnameexp explains the recommendations generated by \pname, and for this experimentation, we \textit{focus} on user-to-city \code{views} relationships, but this works for any relationships. First, we identify an \textit{important} subset of features by measuring the \textit{change in nDCG@K score} when individual features are zeroed out. Then, selecting those that produce a negative impact on nDCG, meaning those features were important for predicting relevant recommendations. To align with the recommendation explainability task, we redefine the traditional Fidelity metric~\cite{yuan2021explainability}—originally based on \textit{change in prediction confidence}—as the \textit{change in nDCG@K score} due to perturbations. The explanation performance of \pnameexp is then evaluated based on the drop in nDCG@K score and the change in cosine similarity ($\Delta \text{sim}$) between user and city embeddings caused by structural perturbations and by only using the \textit{important} features.

\heading{Baseline.} We compare \pnameexp against \gnne~\cite{ying2019gnnexplainer}, \subx~\cite{yuan2021explainability} and PaGE-Link~\cite{zhang2023page}, demonstrating superior explanation quality as evidenced by higher fidelity to the ground truth. We also compare \pname against two baselines to validate its effectiveness: \nm{1} random recommendations where recommendations are generated randomly without considering user preferences, and \nm{2} histogram recommendations where a histogram-based method that creates recommendations from user-item interaction summary. \pname outperforms both the baselines in recommendation accuracy as demonstrated by higher nDCG@K. 

\subsection{Dataset}\label{sec:data}

The dataset was collected over a three-day and 30-day period from the state of Washington in the USA. The dataset statistics are shown in \autoref{tab:summary} and \autoref{tab:summary-edges}, and more details in \autoref{sec:data-stats}. The dataset is split into training, validation, and test sets, ensuring that the test set contains previously unseen pairs, which are evaluated on the following day to avoid any data leakage. The 30-day setting naturally reduces temporal leakage between training and evaluation interactions, further strengthening the case for \pnameexp as a robust, scalable, and domain-aware explainer for real-estate recommender systems. Its ability to combine high fidelity with diversified recommendations makes it the only method that meets practical requirements for accuracy, interpretability, and user engagement.

To note, currently there are no public real estate datasets that contain user-to-listing interactions or the rich attributes we are using. Therefore, we collected and used our own dataset, that accounts for seasonal trends. In the 3-days training data (refer to \autoref{tab:interactions}), expected trends are seen such as view events are the most popular event, followed by save and then toured because users tend to view multiple items before saving them and finally touring the item. The average interaction of the user to listing for view event for the 25th quantile is 1.0 and the 75th quantile is 8.0, the save event for the 25th quantile is 2.7 and the 75th quantile is 3.0, and the tour event for the 25th quantile is 3.2 and the 75th quantile is 4.0. Therefore, from the average interaction of different events statistics tells us that the graph generated from the dataset will contain more sparse connections than dense connections.

The experiments utilize a heterogeneous graph built from user, listing, and city interaction data (shown in \autoref{tab:edge-attr}), where node types encompass multiple attributes of varying data types (shown in \autoref{tab:node-attr}); interested readers can find more details in \autoref{sec:graph-details}. Certain feature values can have valid but outlier values, such as commercial properties with over a thousand bedrooms and bathrooms. In contrast, residential real estate has an average number of bedrooms of 4 and bathrooms of 3.

\begin{figure}[h!]
	\centering
	\includegraphics[width=0.95\linewidth]{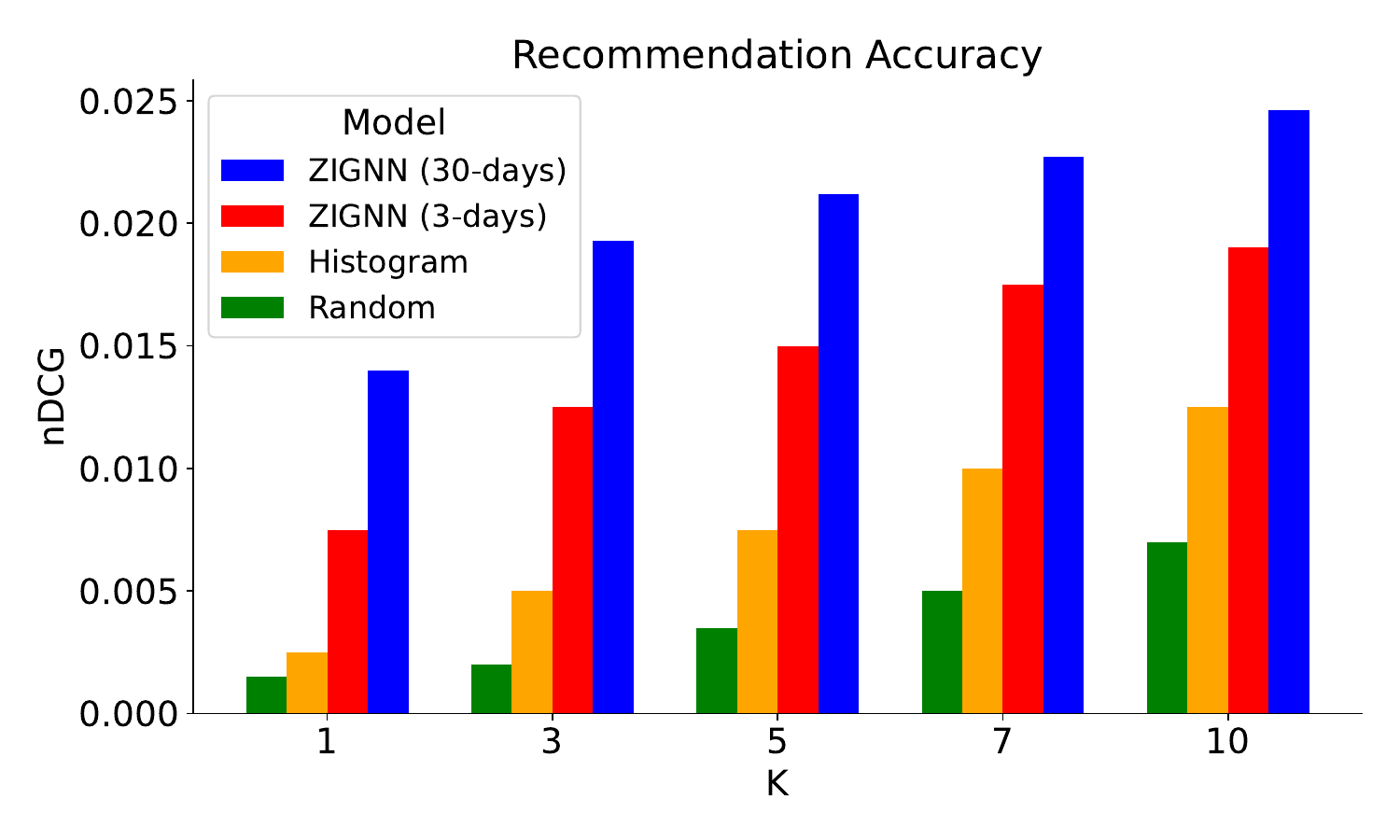}
	\caption{Performance of \pname against different baselines.}
	\label{fig:gnn-perf}
\end{figure}

\subsection{Recommendation Accuracy of \pname}\label{sec:recs-acc}
To investigate \textbf{RQ1}, we compared the performance between the \pname (3-days) and the baseline recommendation models to reveal \pname consistently and substantially outperforms the rest as seen in \autoref{fig:gnn-perf}. \pname consistently surpasses the histogram model across all values of \textit{K}, with the performance gap widening as \textit{K} increases. Notably, at $K=10$, the \pname achieves an nDCG score of 0.019, a 52\% improvement over the histogram model's 0.0125, demonstrating \pname's ability to produce more accurate and relevant recommendations as the recommendation set grows.

This trend is further emphasized by the steep growth curve of the GNN's performance compared to the relatively linear increase observed for the histogram model, showcasing the \pname's adaptability. \pname's structural representation learning captures nuanced relationships in the data that the histogram model, with its simpler statistical approach, fails to address. At $K=1$, where only the top recommendation matters, \pname outperforms the histogram model by a remarkable 200\% (0.0075 vs. 0.0025), highlighting its capacity to prioritize the most relevant results effectively. Comparing \pname performance using the extended 30-day dataset and 3-day dataset, the 30-day model yields consistent improvement gains ranging from ~30\% at $K=10$ to nearly 90\% at $K=1$, highlighting the benefit of incorporating a longer 30-day interaction window where the user behavior is captured with higher relevance.

\begin{figure}[h!]
	\centering
	\includegraphics[width=0.90\linewidth]{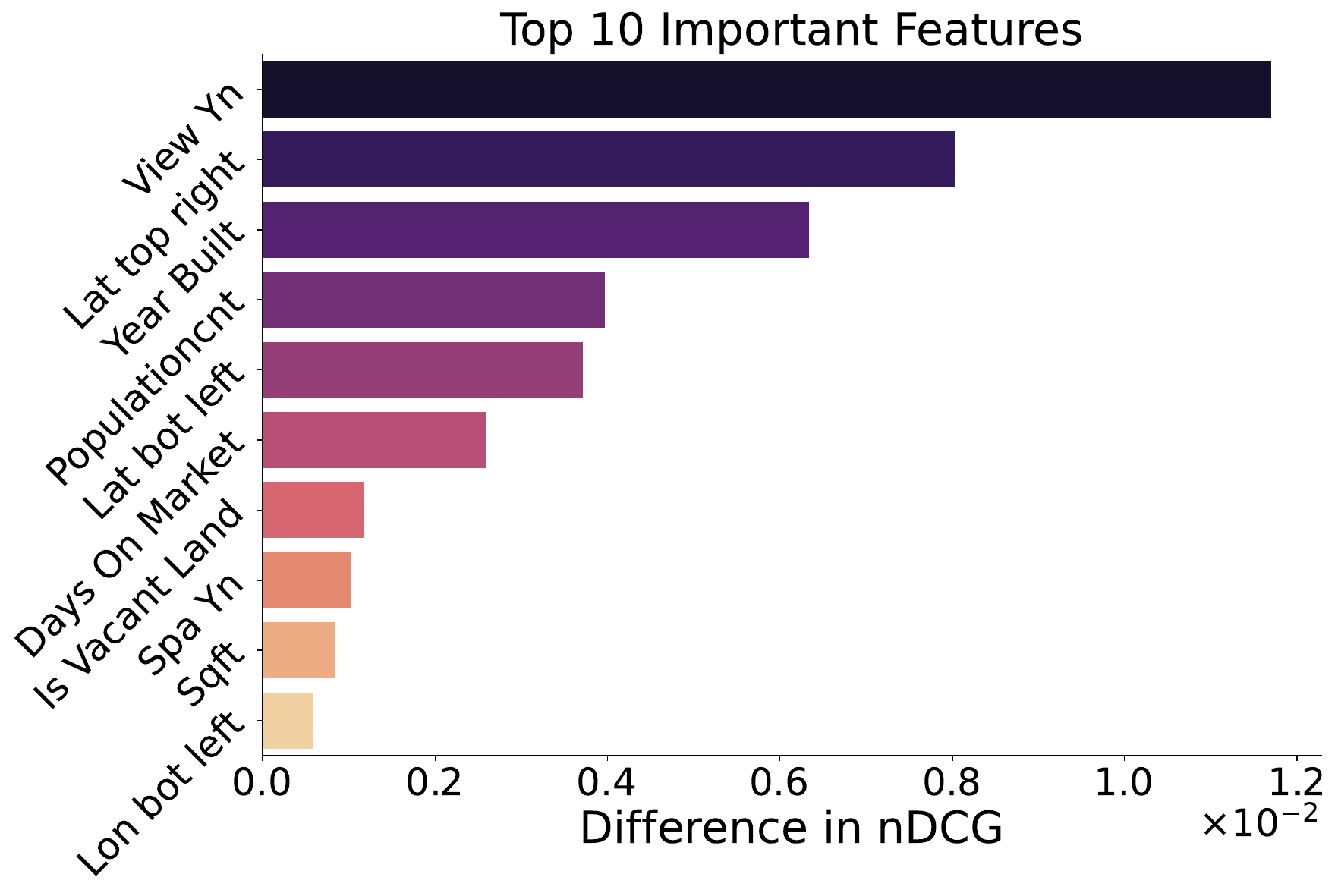}
	\caption{Impact of zeroing out features to find important features.}
	\label{fig:gnn-feat1}
\end{figure}

\begin{figure}[h!]
	\centering
	\includegraphics[width=0.90\linewidth]{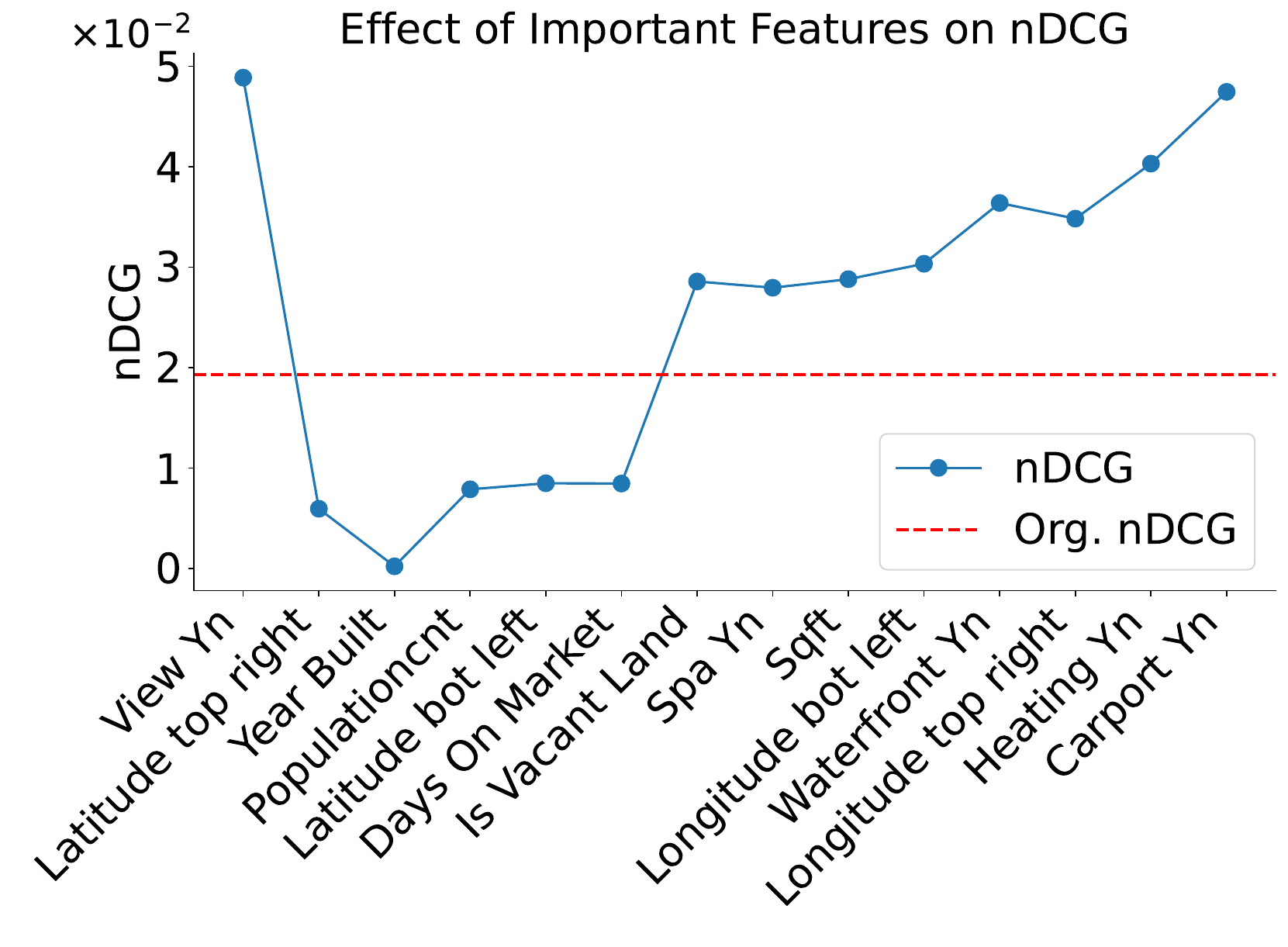}
	\caption{Impact of sequentially adding important features on nDCG.}
	\label{fig:gnn-feat2}
\end{figure}

\subsection{Explanation Quality of \pnameexp}\label{sec:exp-acc}

For investigating \textbf{RQ2}, we first consider the feature perturbation quality and show that we can identify the top ten important city features for the recommendation by measuring the difference between the original nDCG and the perturbed nDCG when a feature is zeroed out. When the difference is positive, the feature was important for predicting the recommendation, as shown in \autoref{fig:gnn-feat1}, and features with negative differences mean they were confusing the model (more details in \autoref{sec:unimp-feat}).

In particular, these important features can be grouped: \nm{1} geographic - latitude (top right and bottom left) and longitude (bottom left), \nm{2} boolean - view, spa, and vacant land, and \nm{3} numeric - year, population count, days on the market, and square feet. As illustrated in \autoref{fig:gnn-feat2}, using only the important features yield better performance than using all the features, demonstrating that \pnameexp{} effectively identifies the important features needed for user-city recommendation. These top features will be the only features used for measuring the structural perturbation quality. 

The important city features align with domain knowledge: a city is typically characterized by its geographic location and aggregate properties (\eg population count, the average year of houses, and average days on the market), which are different from the features usually used to characterize a listing (\eg the number of bathrooms and bedrooms). Aggregate properties offer a holistic snapshot of a city's dynamics and freshness. Interestingly, cities with a younger infrastructure experience rapid growth, indicating a strong interplay between user preference and given recommendations.

In \autoref{fig:gnn-feat1}, the boolean feature \texttt{view} yields the best recommendation performance. Moreover, combining the top fourteen features results in performance similar to the single boolean feature. Incorporating additional features is necessary because relying solely on one boolean feature would compromise the model’s generalizability and robustness. Furthermore, considering a diverse set of features, \pnameexp{} does not overfit its explanation to a single scenario.

\input{tables/expcomp.tex}

\autoref{tab:comparison-metrics} quantitatively evaluates the performance of \pnameexp against \sota \gnn explainers, where \pnameexp outperforms \plink, \gnne, and \subx in both nDCG and cosine similarity metrics, demonstrating its superior explanation accuracy compared with the ground-truth. In the ground-truth, the edges in the real-world dataset are treated as positive, and other edges are treated as negative. It is seen that \pnameexp achieves the most significant reduction in nDCG (94\%), indicating superior fidelity in explanation degradation. Additionally, it achieves the largest decrease in cosine similarity (-0.10), highlighting its effectiveness in finding the important subgraphs that alter \pname's recommendations. 

Our month-long (30-days) evaluation confirms that \pnameexp delivers markedly stronger explanations than all baselines while optimally scaling to a larger interaction graph. With a 30-days training window, \pnameexp preserves 92\% of the original nDCG, almost identical to its 3-day performance (94\%), and maintains the largest embedding shift (-0.09), indicating that it highlights influential edges without collapsing recommendation diversity. 

The stability of \pnameexp across time windows suggests that its structural perturbation strategy generalizes rather than over-fitting to short-term interaction noise. The two-point drop in nDCG retention and the 0.01 improvement in cosine similarity from 3 to 30 days imply that extending the past horizon yields the same explanatory fidelity with slightly more conservative embedding movements, an attractive trade-off for production systems.

\subsection{\sota \gnn Explainers vs. \pnameexp}\label{sec:sota-comp}
We compared \pnameexp against a specialized \gnn-based recommendation explainer \plink~\cite{zhang2023page} and two general \gnn explainers, \gnne~\cite{ying2019gnnexplainer} and \subx~\cite{yuan2021explainability} in \autoref{tab:comparison-metrics} to answer \textbf{RQ3}. \pnameexp performed the best, followed by \plink and general purpose \gnn explainers performed the worst which is expected since \pnameexp and \plink have been specifically designed for \lp tasks. In contrast, the recommendation-specific \plink retains only 63\% of nDCG and induces a much smaller change in cosine similarity, while general-purpose explainers fall to 22\% (\subx) and 7\% (\gnne). Competing methods degrade sharply because their mask-optimization (\gnne) or Monte-Carlo subgraph search (\subx) faces an exponentially expanding candidate space as graph size grows. 

Since, \plink creates the k-hop ego-graph by limiting the consideration of relevant features and subgraphs. Therefore, \plink performance degrades sharply because they compute the k-hop neighborhood around the target link, but there is no guarantee that the k-hop relationship will include relevant links.  This limitation is reflected by -0.04 less decrease is cosine similarity. \pnameexp avoids this pitfall by exploiting domain knowledge: the number of co-clicked cities grows sub-linearly, so the search space for meaningful perturbations remains tractable even on month-long graphs. Compared to general purpose \gnn explainer, \ie \gnne and \subx, \pnameexp demonstrates average of -78\% more reduction in nDCG and -0.08 more cosine similarity increase, indicating that it more effectively identifies the features and subgraphs the \gnn relies on to make recommendations.



\input{tables/tuning.tex}
\subsection{Hyperparameter Sensitivity Study}

\autoref{tab:hparams} shows the hyperparameter sensitivity tuning summary where we test different neighborhood sizes ($K$), edge removal strategy, size of negative graph, learning rate and output feature dimensions. The size of the input feature dimension is governed by the number of features for each entity type, so we did not incorporate that. The three different edge removal strategies we used are: \nm{1} PRI (prioritize Recent Interaction) which removes edges with more recent interactions first (reduces the strength of fresh user-item relationships); \nm{2} HID (High In-Degree) which removes edges to nodes with the highest in-degree (breaking these highly connected links disrupts potential shared neighborhoods); and \nm{3} HBC (High Betweenness Centrality) which removes edges with high betweenness centrality, targeting the links acting as bridges in the hub cities in the interaction graphs.

Increasing the neighborhood size ($K$) from 4 to 8 improves nDCG@1, but going further to 16 diminishes it, indicating that while broader neighborhoods capture more context, too large a neighborhood may dilute relevant signals. HID (high in-degree) as the edge removal strategy yields the best score among the tested strategies, suggesting that selectively removing overly connected edges can help avoid excessive overlap and improve the model’s focus on more distinct interactions. A larger negative graph size, up to a point, seems beneficial (optimal at size 5 in this table) because it gives the model a more balanced signal to distinguish relevant versus irrelevant connections. Tuning the learning rate reveals that a rate of 1e-2 provides a stable convergence path.
Finally, a higher output feature dimension (512) significantly boosts nDCG@1, demonstrating that a richer embedding space helps the model capture more nuanced relational patterns but at the cost of resource consumption.


\begin{figure}[h]
    \centering
    \includegraphics[width=0.70\linewidth]{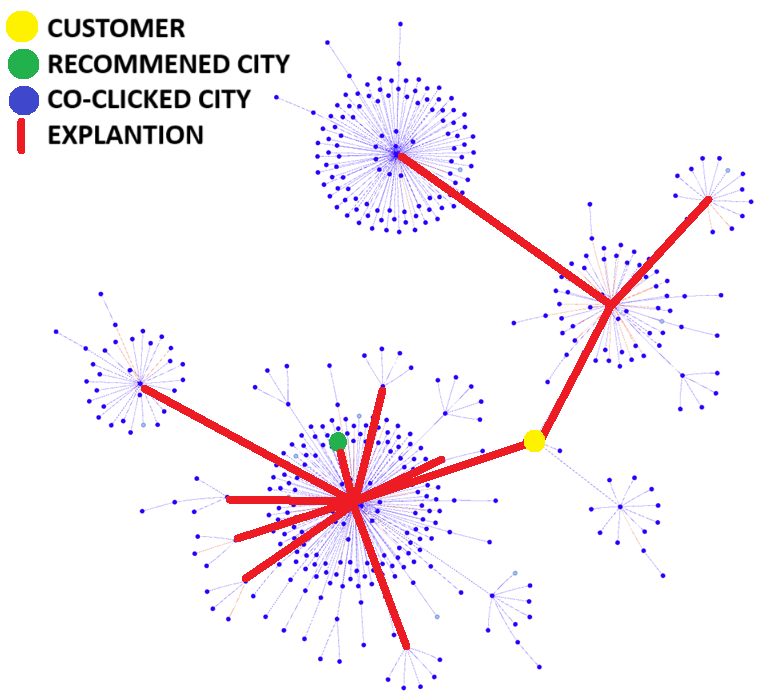}
    \caption{\pnameexp's explanation of a recommended city \#1.}
    \label{fig:exp1}
    \label{fig:sidebyside}
\end{figure}

\section{Case Study}
In the case study, we selected a real user (\ie customer) and, based on their interactions, generated a city recommendation along with explanations.
To enhance visualization, we present a simplified graph (\autoref{fig:exp1}) focusing on the customer (yellow node), co-clicked cities (blue nodes), and the recommended city (green node). A blue edge will exist between the co-clicked city nodes, the recommended city is the green node, and the red edges are the edges identified as important for recommendations. 

\autoref{fig:exp1} illustrates that the user has direct connections to a limited number of co-clicked cities. However, one of these cities is highly connected to other cities, including the recommended city. These highly connected cities act as information hubs, facilitating the flow of information between the user and the recommended city. This information flow occurs through two mechanisms: \nm{1} shared preferences, users who co-click on listings in the hub city exhibit similar preferences to the target user, making the recommended city more relevant, and \nm{2} indirect connections, the hub city connects the user to a broader network of cities with relevant listings, increasing the likelihood of discovering the recommended city. Consequently, \pnameexp identifies the edges connecting the user to these hub cities as critical to the recommendation, as their removal significantly impacts the information flow and potentially disrupts the discovery of relevant recommendations.




%% file: tables/expcomp.tex
\begin{table}[h]
    \centering
    \resizebox{\columnwidth}{!}{%
    \begin{tabular}{@{}lcccc@{}}
    \toprule
    Change in           & \textbf{\pnameexp} & \plink~\cite{zhang2023page} & \gnne~\cite{ying2019gnnexplainer} & \subx~\cite{yuan2021explainability} \\ 
    \midrule
    && 3-days (Training) && \\
    \midrule
    nDCG (\% decr.)     &  \textbf{94\%}    &   81\% (\emph{-13\%})     &     21\% (\emph{-73\%})     &  47\% (\emph{-47\%})       \\
    cosine similarity   & \textbf{-0.10}    &   -0.07 (\emph{-0.03})     &    -0.02 (\emph{-0.08})    &    -0.04 (\emph{-0.06})    \\ 
    \midrule
    && 30-days (Training) && \\
    \midrule
    nDCG (\% decr.)     &  \textbf{92\%}     &   63\% (\emph{-29\%})     &     9\% (\emph{-85\%})     &  22\% (\emph{-70\%})       \\
    cosine similarity   & \textbf{-0.09}    &   -0.05 (\emph{-0.04})     &    -0.01 (\emph{-0.08})    &    -0.02 (\emph{-0.07})    \\ 
    \bottomrule
    \end{tabular}%
    }
    \caption{Quantitative evaluation of \pnameexp against \gnn explainers.}
    \label{tab:comparison-metrics}
    \end{table}

%% file: tables/tuning.tex
\begin{table}[h]
  \centering
  \resizebox{\columnwidth}{!}{%
    \begin{tabular}{@{}llll@{}}
      \toprule
      \textbf{Hyperparameter}        & \textbf{Values Tested}          & \textbf{nDCG@1 Scores}               & \textbf{Best Score}      \\ \midrule
      Neighborhood Size ($k$)        & 4, 8, 16                        & 0.0070, 0.0075, 0.0071               & 0.0075 (at $k{=}8$)      \\
      Edge Removal Strategy          & PRI, HID, HBC                   & 0.0070, 0.0077, 0.0075               & 0.0077 (at \textsc{HID}) \\
      Negative Graph Size         & 1, 3, 5                         & 0.0072, 0.0074, 0.0075               & 0.0075 (at 5)            \\
      Learning Rate                  & $10^{-3}$, $10^{-2}$, $10^{-1}$ & 0.0072, 0.0074, 0.0069               & 0.0074 (at $10^{-2}$)    \\
      Output Feature Dim.       & 128, 256, 512                   & 0.0071, 0.0075, 0.0081               & 0.0081 (at 512)          \\ \bottomrule
    \end{tabular}%
  }
  \caption{Hyperparameter sensitivity summary. Each row varies one hyperparameter while holding the others constant; the rightmost column reports the highest nDCG@1 observed for that setting.}
  \label{tab:hparams}
\end{table}

%% file: sections/conclusion.tex
\section{Conclusion}\label{sec:conclusion}
We introduced \pnameexp, which aims to bridge a critical gap in explainable ML for recommendation systems by providing human-interpretable insights into \gnn-based recommendation systems to increase user confidence and customer engagement. \pnameexp offers contextually relevant and human-interpretable explanations through structural and attribute perturbation. \pnameexp outperforms \plink by 29\%, \gnne by 85\%, and \subx by 70\% in the Fidelity metric, demonstrating its superior explanation accuracy compared with the ground-truth. Specifically, \pname provides whole graph instance-level explanations for \gnn models in the context of the heterogeneous link prediction (LP) task by identifying a subset of node features and subgraphs that are most influential for \gnn's-based recommendation system.


%% file: sections/appendix.tex
\appendix
\section{Appendix}\label{sec:appendix}

\subsection{\pname Architecture}

\pname is a Graph Neural Network-based recommendation engine designed to operate on heterogeneous graphs \( \mathcal{G} \) built using real-world real-estate (\eg Zillow Group, Inc.) interaction data. \pname not only models user-listing (or city) interactions for recommendation but also captures contextual information such as city-level influences and item similarities. The ability to leverage heterogeneous relationships improves the recommendation quality by allowing the system to reason about multi-hop dependencies and indirectly related entities. 

\pname utilized the Zillow Group, Inc. dataset for modeling and evaluation purposes, demonstrating its effectiveness in a real-world recommendation system. However, the architecture of \pname is generic and can be adapted to various platforms across different domains. \pname can be extended to use cases where the context is critical in influencing user decisions, such as (user, item, context/category), where the user and item types will dictate the node types, and the context interaction between the user and item will for the edges. For instance, it can model different interactions: (user, food, restaurant) in the food industry, (user, post, topic) in social media, (patient, doctor, hospital) in healthcare, or (user, job, company) in the job search industry. By leveraging the interaction graph creation and behavioral learning flexibility, \pname can cater to a wide range of recommendation scenarios, enhancing user satisfaction and specific context-aware decision-making across multiple domains.

The \pname consists of the following components:

\begin{enumerate}
    \item Node Embedding Layer: Each node \( v \in \mathcal{V} \) is mapped to a dense vector representation \( \mathbf{h}_v \in \mathbb{R}^d \) using an embedding layer. The initial embeddings are learned from the node features and are iteratively updated during the training process.

    \item Message Passing Mechanism: \pname employs a multi-hop message passing mechanism, where each node \( v \in \mathcal{V} \) aggregates information from its neighbors \( \mathcal{N}(v) \) through a learnable function. The node update rule for \( t \)-th layer is defined as $\mathbf{h}_v^{(t+1)} = \text{AGG}\left( \mathbf{h}_v^{(t)}, \left\{ \mathbf{h}_u^{(t)} : u \in \mathcal{N}(v) \right\} \right)$, where \( \text{AGG} \) is an aggregation function such as sum, mean, or attention-based pooling. This allows the model to capture higher-order dependencies between nodes.

    \item Link Prediction: The final node embeddings \( \mathbf{h}_u \) and \( \mathbf{h}_l \) for users and listings are fed into a scoring function to compute the likelihood of a link using $\hat{y}_{ul} = \sigma(\mathbf{h}_u^T W \mathbf{h}_l)$,    where \( \sigma \) is the sigmoid function and \( W \) is a learnable weight matrix.
\end{enumerate}

\pname is composed of three key components: a projection layer for aligning feature dimensions across different node types, a Relational Graph Convolutional Network (RGCN) for learning node embeddings while considering multi-relational graph structures, and a dot product predictor layer for scoring edge relationships. The projection layer standardizes feature dimensions by applying type-specific linear transformations. This ensures compatibility with the RGCN, which processes node and edge type information through multiple layers of convolution, enabling the model to capture complex interactions between heterogeneous entities.

\begin{figure}[h!]
	\centering
	\includegraphics[width=0.90\linewidth]{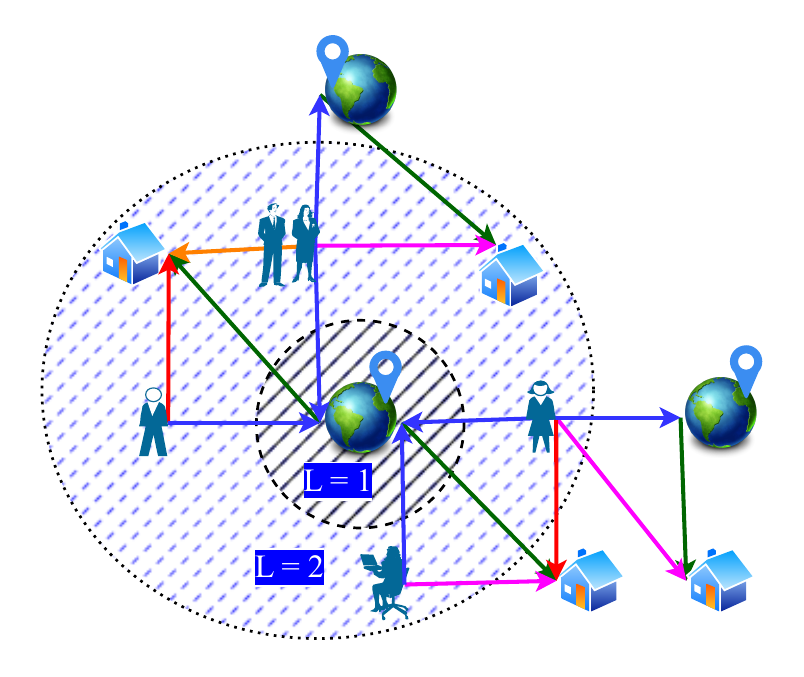}
	\caption{Interaction graph diameter is two so we need at most two GCN layers to incorporate information from the farthest nodes.}
	\label{fig:diameter}
\end{figure}

The RGCN implementation employs a two-layer design, where each layer performs graph convolutions over different relationship types. We purposefully selected two-layer RGCN because the graph diameter of $\mathcal{G}$ is at most two as shown in \autoref{fig:diameter}. Since, the message passing step happens in parallel, the numbers of times the message passing steps need to be completed for the information to travel from the farthest part of the graph is two.

For each relationship, the model constructs a separate Graph Convolutional layer and applies type-specific linear transformations to incorporate edge-specific features into the convolution process. Additionally, self-loop embeddings are refined using residual connections for each node type, ensuring that the node's initial features are preserved alongside learned representations. This structure allows the RGCN to aggregate information across the graph, dynamically updating node embeddings while addressing the unique characteristics of heterogeneous relationships.

The dot product predictor layer scores edges by computing the dot product between node embeddings at the source and target ends of an edge for each edge type. During this process, the predictor assigns scores to edges by using the learned node features (h) from the RGCN, which encapsulate the structural and relational context of each node in the graph. For positive edges, which represent actual relationships observed in the graph, the scores reflect the strength of these connections as encoded in the node embeddings. Conversely, for negative edges, which are artificially sampled to represent non-existent or unlikely relationships, the scores typically indicate weaker or negligible associations. This distinction between positive and negative edge scores is critical for the model to learn meaningful embeddings and effectively differentiate between real and spurious relationships, forming the basis for accurate recommendation tasks.

The implementation is built using \dgl for graph operations and PyTorch for neural network components. FAISS is integrated to perform efficient nearest-neighbor searches for recommendation evaluation. 

During training, both positive and negative edge scores are calculated for each edge type by applying the dot-product predictor to pairs of node embeddings, enabling the model to distinguish between observed and unobserved relationships. We use the Adam optimizer with weighted decay to optimize margin-based loss, where the positive edge scores are encouraged to exceed negative edge scores by at least a margin of 1, penalizing cases where this condition is not met. The modular design supports inference by directly returning node embeddings when negative edges are not provided. This architecture ensures the model can efficiently learn and generalize from multi-relational graph data for edge classification tasks.

\input{algorithm/zrex-algo.tex}

\input{algorithm/postprocessing.tex}

\subsection{\pnameexp Algorithm}

\autoref{alg:zrex} describes the \pnameexp algorithm and \autoref{alg:preprocess} describes the algorithm to generate the heterogeneous interaction graph, $G_{hetero}$ from clickstream events. \autoref{alg:preprocess} converting raw interaction and regional metadata CSVs into attribute-normalized heterogeneous graph, $G_{hetero}$, that connects the users, listings, and cities through user-listing, user-city, and city-listing edges. 

Using $G_{hetero}$, \autoref{alg:zrex} delivers fine-grained model transparency: for a given user-city recommendation, it \nm{1} ranks city-level features by zeroing them out and measuring the resulting $\Delta nDCG@K$, and \nm{2} ranks graph edges in the $K$-hop neighbourhood by perturbing edge and observing the change in the cosine similarity between the \pname embeddings of the user and target city. The joint output—two ordered lists of high-impact features and structural subgraphs, exposes the concrete evidence that the GNN relied upon.

\input{tables/attr.tex}

\subsection{Node and Edge Details}\label{sec:graph-details}

As described in the preliminaries \autoref{sec:inter-graph}, there are three different types of nodes: user, listing, and city. There are different relationships or edges between them, as shown in \autoref{tab:edge-attr}. The most popular relationship between the user and the listing is the \code{views} relationship, since users view multiple listings before narrowing down their search by saving the listing and finally touring the listing. Between user and city, \code{searched in} relationship exists, and between city and listing \code{contains} relationship exists. 

\autoref{tab:node-attr} outlines the attribute schema for the different node types used in our model. The User node is characterized by a numeric session identifier. The Listing node includes a diverse set of features: numeric attributes (\eg bedrooms, bathrooms, price, etc.), boolean attributes (\eg waterfront, heating, etc.), and detailed geographic attributes (latitude and longitude for the property boundaries). Meanwhile, the city node aggregates listing data by averaging numeric, boolean, and geographic attributes and adding population count. 

\input{tables/dataset.tex}
\input{tables/dataset-edges.tex}
\input{tables/interaction.tex}

\subsection{Dataset Statistics}\label{sec:data-stats}

The dataset statistics as shown in \autoref{tab:summary} and \autoref{tab:summary-edges} shows that the dataset is divided into four segments: 3-day training (May 17-20), month-long training (April 20 - May 20), testing (May 27-30), and evaluation (May 31)—providing a temporal snapshot of various metrics.  The 3-day training dataset, spanning May 17th to May 20th, includes 55k listings viewed by 393k users across 449 cities, resulting in 789k \code{views}, 169k \code{saves}, and 234k \code{tours}. The \code{contains} relationship matches the number of listings since each listing must belong to a city. The testing dataset exhibits similar characteristics from May 27th to May 30th with slightly different values (53k listings and 405k users). Notably, the evaluation dataset on May 31st, while covering a similar number of cities (448), shows a smaller number of listings (45k), users (203k), saves (41k), and tours (56k) in comparison to training and testing datasets since there is a decrease in user engagement during the evaluation period which is just one day after the testing period. The \code{views} are the largest interactions since users view the listing the most, and only a few of them convert to \code{saves}, and fewer convert to \code{tours}.

As illustrated in \autoref{tab:interactions}, user engagement with listings scales markedly with the observation window. Over the 7-day horizon, users perform a mean of 7.97 views, 2.74 saves, and 3.24 tour requests per listing, yet the corresponding medians (3, 2, 2) reveal a highly right-skewed distribution in which a small subset of users interactions form the majority of the engagement dataset. Extending the window to 30 days amplifies this effect: mean interactions roughly triple for views and more than double for both saves and tours, while the medians increase more modestly (to 5, 2, 3), indicating that additional time primarily benefits heavier-engagement cohorts. The widening gap between the 25\textsuperscript{th} and 75\textsuperscript{th} percentiles across all metrics further shows the importance of the growing dispersion, suggesting that long-term recommendation strategies should explicitly account for this heterogeneity rather than relying on aggregate averages alone.

\begin{figure}[h!]
	\centering
	\includegraphics[width=\linewidth]{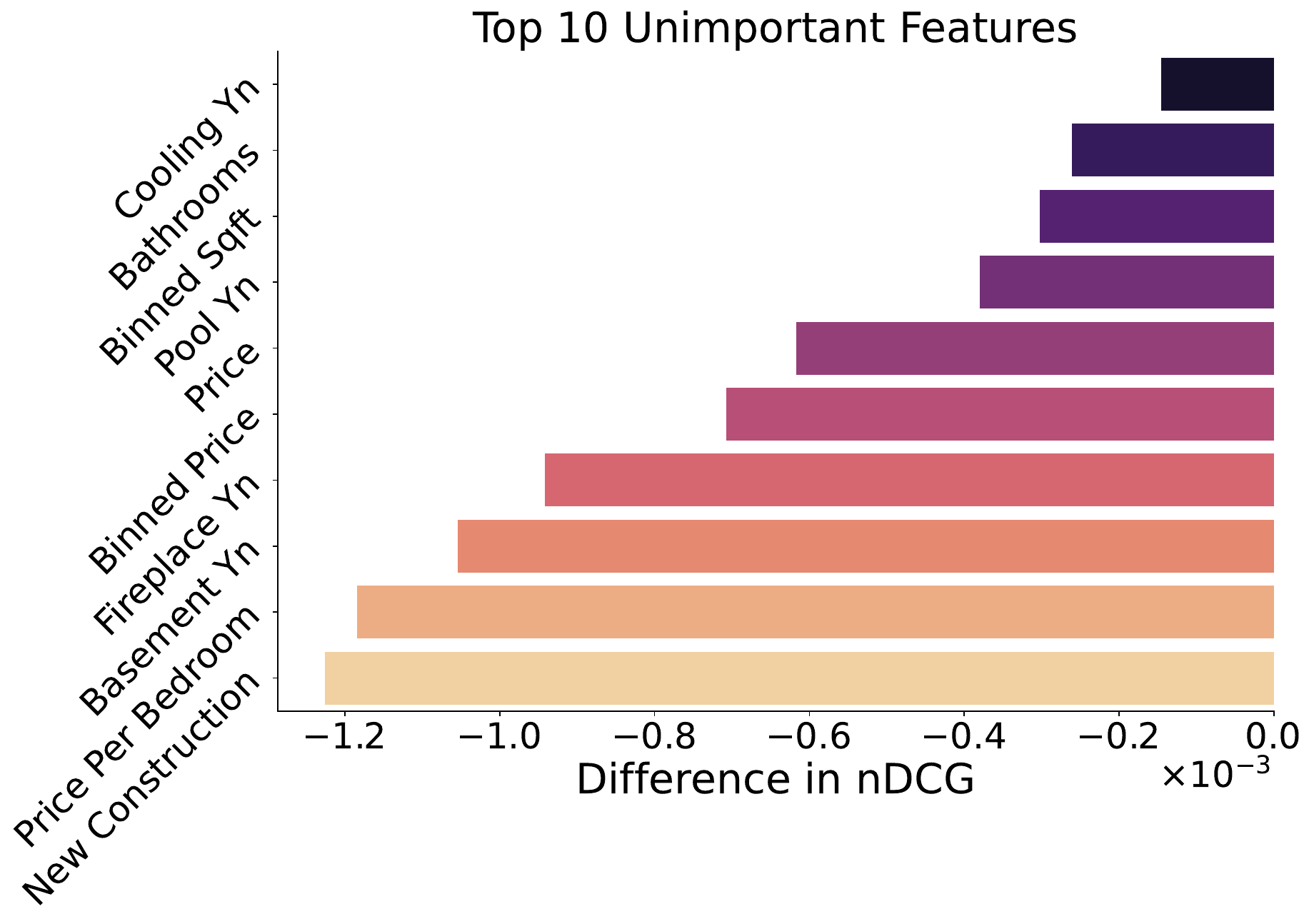}
	\caption{Impact of zeroing out features to find unimportant features.}
	\label{fig:gnn-feat3}
\end{figure}

\subsection{Unimportant Features}\label{sec:unimp-feat}

The unimportant features, as shown in \autoref{fig:gnn-feat3}, are identified by measuring the performance increase when the features are zeroed out, indicating that the feature was detrimental to the \pname’s predictive performance. Therefore, these are unimportant features and should be removed from the model to enhance accuracy. It is interesting to see that the numeric unimportant features are usually used to identify listing. Understandably, aggregating those listing features is not helpful to \pname when compared against city-specific features, such as population count, average year built, and geographic-based features. 

\input{tables/time-overhead}
\subsection{Runtime Analysis: 3 vs 30 Day Dataset}
\autoref{tab:runtime} shows that enlarging the interaction window from 3 days (1.2M edges) to 30 days (29.8M edges) raises the cost of one training epoch from 0.88s to only 10.89s.  Despite a 25 times increase in graph size, the runtime grows by just one order of magnitude, confirming that our training pipeline scales sub‑linearly with temporal coverage.  Inference remains virtually size‑independent: generating an explanation for a single user takes 5.14s on the 3‑day graph versus 5.04s on the 30‑day graph.  Together, these results demonstrate that \pname delivers timely and accurate explanations at production scale.

\balance

%% file: algorithm/zrex-algo.tex
\begin{algorithm}[t]
\caption{Z‑REx}
\label{alg:zrex}
\begin{algorithmic}[1]
\Require target user $u$, target city $c_t$, heterogeneous interaction graph $G_{\text{hetero}}$, trained GNN \textsf{ZiGNN}, rank cut‑off $K$, city‑feature $F_{\text{all}}$, hop distance $k$

\Function{Z‑REx}{$u,c_t,F_{\text{all}},k$}
    \State $\textit{ranked\_features} \gets \Call{FeaturePerturb}{u,c_t,F_{\text{all}}}$
    \State $\textit{ranked\_edges}    \gets \Call{StructuralPerturb}{u,c_t,k}$
    \State \Return $\textit{ranked\_features}, \textit{ranked\_edges}$
\EndFunction

\vspace{0.3em}
\Function{FeaturePerturb}{$u,c_t,F_{\text{all}}$}
    \State $baseline\_ndcg \gets \Call{NDCG}{u}$  \Comment original score
    \State $\textit{ranked\_features} \gets \{\}$
    \ForAll{$f \in F_{\text{all}}$}
        \State $\bar{x}_{c_t} \gets \Call{ZeroOutFeature}{c_t,f}$
        \State $\Delta ndcg \gets \Call{NDCG}{u,\bar{x}_{c_t}} - baseline\_ndcg$ \Comment Eq.(3)
        \State $\textit{ranked\_features} \gets \textit{ranked\_features} \cup (f,\Delta ndcg)$
    \EndFor
    \State \Call{Sort}{$\textit{ranked\_features}$} by $\Delta ndcg$ (descending)
    \State \Return $\textit{ranked\_features}$
\EndFunction

\vspace{0.3em}
\Function{StructurePerturb}{$u,c_t,k$}
    \State $G_h \gets \Call{CollapseToUserCityGraph}{G_{\text{hetero}}, c_t}$
    \State $G_k \gets \Call{KHopSubgraph}{G_h,u,k}$
    \State \Call{AddCo-ClickEdges}{$G_k$}
    \State $base\_sim \gets \text{cos}(\textsf{ZiGNN.embed}(u),\textsf{ZiGNN.embed}(c_t))$
    \State $\textit{ranked\_edges} \gets \{\}$
    \ForAll{edge $e$ incident to $u$ or $c_t$ in $G_k$}
        \State temporarily remove $e$
        \State $new\_sim \gets \text{cos}(\textsf{ZiGNN.embed}(u),\textsf{ZiGNN.embed}(c_t))$
        \State $\Delta sim \gets new\_sim - base\_sim$ \Comment Eq.(4)
        \State $\textit{ranked\_edges} \gets \textit{ranked\_edges} \cup (e,|\Delta sim|)$
        \State restore $e$
    \EndFor
    \State \Call{Sort}{$\textit{ranked\_edges}$} by $|\Delta sim|$ (descending)
    \State \Return $\textit{ranked\_edges}$
\EndFunction

\end{algorithmic}
\end{algorithm}

%% file: algorithm/postprocessing.tex
\begin{algorithm}[t]
\caption{Generate Heterogeneous Interaction Graph}
\label{alg:preprocess}
\begin{algorithmic}[1]
\Require CSVs (\ie $events\_file$, $regions\_file$), normalization method $norm\_method$

\Function{Preprocess}{$events\_file, regions\_file$}
    \State $events  \gets \Call{ReadCSV}{events\_file}$
    \State $regions \gets \Call{ReadCSV}{regions\_file}$
    \vspace{0.3em}
    
    \State $U_L \gets \Call{BuildUserListingEdges}{events}$
    \State $regions' \gets regions[id \in U_L.listing\_id]$

    \State $U_C \gets \Call{BuildUserCityEdges}{U_L,regions'}$
    \State $C_L \gets \Call{BuildCityListingEdges}{regions'}$
    \vspace{0.3em}

    \State $users    \gets \Call{Unique}{events.user\_id, Xavier\_Init}$
    \State $listings \gets \Call{CleanNumeric}{regions', norm\_method}$
    \State $cities   \gets \Call{AggregateCity}{regions', norm\_method}$

    \State $G_{hetero}      \gets \Call{dgl.heterograph}{U_L,U_C,C_L}$

    \State \Return $G_{hetero}$
\EndFunction
\end{algorithmic}
\end{algorithm}

%% file: tables/attr.tex
\begin{table}[ht]
    \centering
    \resizebox{\columnwidth}{!}{%
    \begin{tabular}{@{}lcc@{}}
    \toprule
    \textbf{\makecell{Node \\ Type}} & \textbf{Attribute} & \textbf{\makecell{Attribute \\ Type}} \\ \midrule
    User               & session id                & Numeric                 \\ \addlinespace
    \cline{2-3}
    \addlinespace
    Listing            & \makecell{bedrooms, bathrooms, year built, sq. ft, price \\ binned sq. ft, binned price, price per bedroom \\ days on market, floors}           & Numeric                 \\
    &  & \\
    Listing            & \makecell{Waterfront, heating, basement, fireplace \\ Cooling, view, vacant, spa, carport \\ Pool, new construction}              & Boolean                 \\
    &  & \\
    Listing            & \makecell{Latitude top/bottom left/ right \\ Longitude top/bottom left/right}            & Geographic              \\ 
    \addlinespace
    \cline{2-3}
    \addlinespace
    City               & \makecell{population count, avg. all the listing features \\ for listings per city }  & Numeric                 \\
    \addlinespace
    City               & \makecell{avg. all the boolean features \\ for listings per city }  & Boolean                 \\
    \addlinespace
    City               & \makecell{avg. all the geographic features \\ for listings per city }  & Geographic                 \\
    \bottomrule
    \end{tabular}%
    }
    \caption{Attributes and their types for different node types.}
    \label{tab:node-attr}
\end{table}

%% file: tables/dataset.tex
\begin{table}[ht]
        \centering
        \resizebox{0.8\columnwidth}{!}{%
        \begin{tabular}{@{}lcccc@{}}
        \toprule
                 & \textbf{Date}    & \textbf{City} & \textbf{Listing} & \textbf{User} \\ \midrule
        3-days Training              & 5/17-5/20     & 449          & 55k             & 393k                        \\
        30-days Training         & 4/20-5/20     & 456          & 97k             & 1.3M                        \\
        Testing                     & 5/27-5/30     & 452          & 53k             & 405k                        \\
        Evaluation                  & 5/31          & 448          & 45k             & 203k                        \\ \bottomrule
        \end{tabular}%
        }
        \caption{Number of entities (\eg nodes) in different datasets.}
        \label{tab:summary}
        \end{table}

%% file: tables/dataset-edges.tex
\begin{table}[ht]
        \centering
        \resizebox{0.9\columnwidth}{!}{%
        \begin{tabular}{@{}lccccc@{}}
        \toprule
                 & \textbf{Date}    & \textbf{\code{views}}  & \textbf{\code{saves}} & \textbf{\code{tours}} & \textbf{\code{contains}} \\ \midrule
        3-days Training              & 5/17-5/20              & 789k               & 169k               & 234k               & 55k               \\
        30-days Training         & 4/20-5/20              & 26M                & 2.2M               & 1.6M               & 97k               \\
        Testing                     & 5/27-5/30              & 773k               & 138k               & 197k               & 53k               \\
        Evaluation                  & 5/31                   & 714k               & 41k                & 56k                & 45k               \\ \bottomrule
        \end{tabular}%
        }
        \caption{Number of relationships (\eg edges) in different datasets.}
        \label{tab:summary-edges}
        \end{table}

%% file: tables/interaction.tex
\begin{table}[ht]
  \centering
  \resizebox{0.65\columnwidth}{!}{%
    \begin{tabular}{lccc}
      \toprule
      & \code{views} & \code{saves} & \code{tours} \\
      \midrule
      & & 3-days & \\
      \cline{2-4}  \\
      mean          & 7.97 & 2.74 & 3.24 \\
      25\textsuperscript{th} quantile & 1.00 & 1.00 & 1.00 \\
      median        & 3.00 & 2.00 & 2.00 \\
      75\textsuperscript{th} quantile & 8.00 & 3.00 & 4.00 \\
      \midrule
      & & 30-days & \\
      \cline{2-4}  \\
      mean          & 21.90 & 5.65 & 6.72 \\
      25\textsuperscript{th} quantile & 2.00 & 1.00 & 1.00 \\
      median        & 5.00 & 2.00 & 3.00 \\
      75\textsuperscript{th} quantile & 18.00 & 6.00 & 7.00 \\
      \bottomrule
    \end{tabular}
  }
  \caption{Average number of user-city interactions.}
  \label{tab:interactions}
\end{table}

%% file: tables/time-overhead.tex
\begin{table}[h]
  \centering
  \caption{\pname Runtime performance on different dataset windows.}
  \label{tab:runtime}
  \resizebox{0.6\columnwidth}{!}{%
    \begin{tabular}{@{}lccc@{}}
      \toprule
      & \textbf{Scenario} & \textbf{\# of Edges} & \textbf{Time (s)} \\ 
      \midrule
      \multicolumn{4}{c}{3‑days} \\
      \midrule
        Training & per epoch  & 1.2M  & 0.88 \\
        Evaluation       & per user   & 248k  & 5.14 \\
      \midrule
      \multicolumn{4}{c}{30‑days} \\
      \midrule
      Training  & per epoch  & 29.8M & 10.89  \\
      Evaluation       & per user   & 248k  & 5.04 \\
      \bottomrule
    \end{tabular}%
  }
\end{table}

%% file: refs/gnnrecs.bib
@misc{dgl,
  title        = {DEEP GRAPH LIBRARY: Easy Deep Learning on Graphs},
  howpublished = {\url{https://www.dgl.ai/}},
    year = 2019
}

@misc{zillowwebsite,
  howpublished = {\url{https:www.zillow.com}},
  title        = {Zillow Group, Inc website}
}

@inproceedings{ying2019gnnexplainer,
  author    = {Ying, Zhitao and Bourgeois, Dylan and You, Jiaxuan and Zitnik, Marinka and Leskovec, Jure},
  booktitle = {Neural Information Processing Systems},
  title     = {Gnnexplainer: Generating explanations for graph neural networks},
  year      = {2019},
  ids       = {gnnexplainer}
}

@inproceedings{luo2020parameterized,
  ids     = {pgexplainer},
  author  = {Luo, Dongsheng and Cheng, Wei and Xu, Dongkuan and Yu, Wenchao and Zong, Bo and Chen, Haifeng and Zhang, Xiang},
  booktitle = {Neural Information Processing Systems},
  title   = {Parameterized explainer for graph neural network},
  year    = {2020}
}

@inproceedings{yuan2021explainability,
  ids          = {subgraphx},
  author       = {Yuan, Hao and Yu, Haiyang and Wang, Jie and Li, Kang and Ji, Shuiwang},
  booktitle    = {International Conference on Machine Learning},
  title        = {On explainability of graph neural networks via subgraph explorations},
  year         = {2021},
  organization = {PMLR}
}

@inproceedings{zhang2023page,
  title={PaGE-Link: Path-based graph neural network explanation for heterogeneous link prediction},
  author={Zhang, Shichang and Zhang, Jiani and Song, Xiang and Adeshina, Soji and Zheng, Da and Faloutsos, Christos and Sun, Yizhou},
  booktitle={Proceedings of the ACM Web Conference 2023},
  pages={3784--3793},
  year={2023}
}

@inproceedings{chen2020revisiting,
  title={Revisiting graph based collaborative filtering: A linear residual graph convolutional network approach},
  author={Chen, Lei and Wu, Le and Hong, Richang and Zhang, Kun and Wang, Meng},
  booktitle={Proceedings of the AAAI conference on artificial intelligence},
  volume={34},
  number={01},
  pages={27--34},
  year={2020}
}

@inproceedings{he2020lightgcn,
  title={Lightgcn: Simplifying and powering graph convolution network for recommendation},
  author={He, Xiangnan and Deng, Kuan and Wang, Xiang and Li, Yan and Zhang, Yongdong and Wang, Meng},
  booktitle={Proceedings of the 43rd International ACM SIGIR conference on research and development in Information Retrieval},
  pages={639--648},
  year={2020}
}

@inproceedings{li2019hierarchical,
  title={Hierarchical Representation Learning for Bipartite Graphs.},
  author={Li, Chong and Jia, Kunyang and Shen, Dan and Shi, C-J Richard and Yang, Hongxia},
  booktitle={IJCAI},
  volume={19},
  pages={2873--2879},
  year={2019}
}

@inproceedings{sun2020neighbor,
  title={Neighbor interaction aware graph convolution networks for recommendation},
  author={Sun, Jianing and Zhang, Yingxue and Guo, Wei and Guo, Huifeng and Tang, Ruiming and He, Xiuqiang and Ma, Chen and Coates, Mark},
  booktitle={Proceedings of the 43rd international ACM SIGIR conference on research and development in information retrieval},
  pages={1289--1298},
  year={2020}
}

@inproceedings{wang2019neural,
  title={Neural graph collaborative filtering},
  author={Wang, Xiang and He, Xiangnan and Wang, Meng and Feng, Fuli and Chua, Tat-Seng},
  booktitle={Proceedings of the 42nd international ACM SIGIR conference on Research and development in Information Retrieval},
  pages={165--174},
  year={2019}
}

@inproceedings{sun2019multi,
  title={Multi-graph convolution collaborative filtering},
  author={Sun, Jianing and Zhang, Yingxue and Ma, Chen and Coates, Mark and Guo, Huifeng and Tang, Ruiming and He, Xiuqiang},
  booktitle={2019 IEEE International Conference on Data Mining (ICDM)},
  pages={1306--1311},
  year={2019},
  organization={IEEE}
}

@article{liu2020deoscillated,
  title={Deoscillated graph collaborative filtering},
  author={Liu, Zhiwei and Meng, Lin and Jiang, Fei and Zhang, Jiawei and Yu, Philip S},
  journal={arXiv preprint 2011.02100},
  year={2020}
}

@article{hamilton2017inductive,
  title={Inductive representation learning on large graphs},
  author={Hamilton, Will and Ying, Zhitao and Leskovec, Jure},
  journal={Advances in neural information processing systems},
  volume={30},
  year={2017}
}

@inproceedings{ying2018graph,
  title={Graph convolutional neural networks for web-scale recommender systems},
  author={Ying, Rex and He, Ruining and Chen, Kaifeng and Eksombatchai, Pong and Hamilton, William L and Leskovec, Jure},
  booktitle={Proceedings of the 24th ACM SIGKDD international conference on knowledge discovery \& data mining},
  pages={974--983},
  year={2018}
}

@article{wu2022graph,
  title={Graph neural networks in recommender systems: a survey},
  author={Wu, Shiwen and Sun, Fei and Zhang, Wentao and Xie, Xu and Cui, Bin},
  journal={ACM Computing Surveys},
  volume={55},
  number={5},
  pages={1--37},
  year={2022},
  publisher={ACM New York, NY}
}

@article{zhang2020explainable,
  title={Explainable recommendation: A survey and new perspectives},
  author={Zhang, Yongfeng and Chen, Xu and others},
  journal={Foundations and Trends{\textregistered} in Information Retrieval},
  volume={14},
  number={1},
  pages={1--101},
  year={2020},
  publisher={Now Publishers, Inc.}
}

@article{lyu2022knowledge,
  title={Knowledge enhanced graph neural networks for explainable recommendation},
  author={Lyu, Ziyu and Wu, Yue and Lai, Junjie and Yang, Min and Li, Chengming and Zhou, Wei},
  journal={IEEE Transactions on Knowledge and Data Engineering},
  volume={35},
  number={5},
  pages={4954--4968},
  year={2022},
  publisher={IEEE}
}

@inproceedings{sinha2002role,
  title={The role of transparency in recommender systems},
  author={Sinha, Rashmi and Swearingen, Kirsten},
  booktitle={CHI'02 extended abstracts on Human factors in computing systems},
  pages={830--831},
  year={2002}
}

@inproceedings{wang2022multi,
  title={Multi-level recommendation reasoning over knowledge graphs with reinforcement learning},
  author={Wang, Xiting and Liu, Kunpeng and Wang, Dongjie and Wu, Le and Fu, Yanjie and Xie, Xing},
  booktitle={Proceedings of the ACM Web Conference 2022},
  pages={2098--2108},
  year={2022}
}

@inproceedings{zhang2014explicit,
  title={Explicit factor models for explainable recommendation based on phrase-level sentiment analysis},
  author={Zhang, Yongfeng and Lai, Guokun and Zhang, Min and Zhang, Yi and Liu, Yiqun and Ma, Shaoping},
  booktitle={Proceedings of the 37th international ACM SIGIR conference on Research \& development in information retrieval},
  pages={83--92},
  year={2014}
}

@inproceedings{he2015trirank,
  title={Trirank: Review-aware explainable recommendation by modeling aspects},
  author={He, Xiangnan and Chen, Tao and Kan, Min-Yen and Chen, Xiao},
  booktitle={Proceedings of the 24th ACM international on conference on information and knowledge management},
  year={2015}
}

@inproceedings{peake2018explanation,
  title={Explanation mining: Post hoc interpretability of latent factor models for recommendation systems},
  author={Peake, Georgina and Wang, Jun},
  booktitle={Proceedings of the 24th ACM SIGKDD International Conference on Knowledge Discovery \& Data Mining},
  pages={2060--2069},
  year={2018}
}

@inproceedings{duval2021graphsvx,
  title={Graphsvx: Shapley value explanations for graph neural networks},
  author={Duval, Alexandre and Malliaros, Fragkiskos D},
  booktitle={Machine Learning and Knowledge Discovery in Databases. Research Track: European Conference, ECML PKDD 2021, Bilbao, Spain, September 13--17, 2021, Proceedings, Part II 21},
  pages={302--318},
  year={2021},
  organization={Springer}
}

@inproceedings{lin2021generative,
  title={Generative causal explanations for graph neural networks},
  author={Lin, Wanyu and Lan, Hao and Li, Baochun},
  booktitle={International Conference on Machine Learning},
  pages={6666--6679},
  year={2021},
  organization={PMLR}
}

@article{lundberg2017unified,
  title={A unified approach to interpreting model predictions},
  author={Lundberg, Scott},
  journal={arXiv preprint arXiv:1705.07874},
  year={2017}
}

@inproceedings{ribeiro2016should,
  title={" Why should i trust you?" Explaining the predictions of any classifier},
  author={Ribeiro, Marco Tulio and Singh, Sameer and Guestrin, Carlos},
  booktitle={Proceedings of the 22nd ACM SIGKDD international conference on knowledge discovery and data mining},
  pages={1135--1144},
  year={2016}
}

@inproceedings{pope2019explainability,
  title={Explainability methods for graph convolutional neural networks},
  author={Pope, Phillip E and Kolouri, Soheil and Rostami, Mohammad and Martin, Charles E and Hoffmann, Heiko},
  booktitle={Proceedings of the IEEE/CVF conference on computer vision and pattern recognition},
  year={2019}
}

@inproceedings{harrison2022taxonomic,
  title={Taxonomic Recommendations of Real Estate Properties with Textual Attribute Information},
  author={Harrison, Zachary and Khazane, Anish},
  booktitle={Proceedings of the 16th ACM Conference on Recommender Systems},
  pages={479--481},
  year={2022}
}

@inproceedings{wang2023collaboration,
  title={Collaboration-aware graph convolutional network for recommender systems},
  author={Wang, Yu and Zhao, Yuying and Zhang, Yi and Derr, Tyler},
  booktitle={Proceedings of the ACM web conference},
  year={2023}
}

@article{mukherjee2023explaining,
  title={Explaining Provenance-Based GNN Detectors with Graph Structural Features},
  author={Mukherjee, Kunal and Wiedemeier, Joshua and Wang, Tianhao and Kim, Muhyun and Chen, Feng and Kantarcioglu, Murat and Jee, Kangkook},
  journal={arXiv preprint arXiv:2306.00934},
  year={2023}
}
